\documentclass[12pt,british,twoside]{article}
\usepackage[T1]{fontenc}
\usepackage[latin9]{inputenc}
\usepackage[a4paper]{geometry}
\usepackage{amsmath}
\usepackage{amssymb}
\usepackage{setspace}
\makeatletter
\newcommand{\lyxaddress}[1]{
\par {\raggedright #1
\vspace{1.4em}
\noindent\par}
}

\usepackage{graphicx}
\usepackage{xcolor}
\definecolor{darkblue}{rgb}{0,0,0.5} 
\usepackage{transparent}
\usepackage{microtype}
\newcommand{\fermionpropright}{%% Creator: Inkscape inkscape 0.91+devel+osxmenu, www.inkscape.org
\begingroup%
  \makeatletter%
  \providecommand\color[2][]{%
    \errmessage{(Inkscape) Color is used for the text in Inkscape, but the package 'color.sty' is not loaded}%
    \renewcommand\color[2][]{}%
  }%
  \providecommand\transparent[1]{%
    \errmessage{(Inkscape) Transparency is used (non-zero) for the text in Inkscape, but the package 'transparent.sty' is not loaded}%
    \renewcommand\transparent[1]{}%
  }%
  \providecommand\rotatebox[2]{#2}%
  \ifx\svgwidth\undefined%
    \setlength{\unitlength}{75.00000169bp}%
    \ifx\svgscale\undefined%
      \relax%
    \else%
      \setlength{\unitlength}{\unitlength * \real{\svgscale}}%
    \fi%
  \else%
    \setlength{\unitlength}{\svgwidth}%
  \fi%
  \global\let\svgwidth\undefined%
  \global\let\svgscale\undefined%
  \makeatother%
  \begin{picture}(1,0.32836877)%
    \put(0,0){\includegraphics[width=\unitlength,page=1]{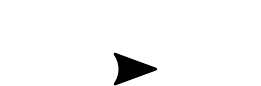}}%
    \put(0.50048819,0.22498336){\color[rgb]{0,0,0}\makebox(0,0)[b]{\smash{$k\,,ba$}}}%
    \put(0,0){\includegraphics[width=\unitlength,page=2]{Fermion_Propagator_Right.pdf}}%
  \end{picture}%
\endgroup%
}
\newcommand{\fermionpropleft}{%% Creator: Inkscape inkscape 0.91+devel+osxmenu, www.inkscape.org
\begingroup%
  \makeatletter%
  \providecommand\color[2][]{%
    \errmessage{(Inkscape) Color is used for the text in Inkscape, but the package 'color.sty' is not loaded}%
    \renewcommand\color[2][]{}%
  }%
  \providecommand\transparent[1]{%
    \errmessage{(Inkscape) Transparency is used (non-zero) for the text in Inkscape, but the package 'transparent.sty' is not loaded}%
    \renewcommand\transparent[1]{}%
  }%
  \providecommand\rotatebox[2]{#2}%
  \ifx\svgwidth\undefined%
    \setlength{\unitlength}{75.00000169bp}%
    \ifx\svgscale\undefined%
      \relax%
    \else%
      \setlength{\unitlength}{\unitlength * \real{\svgscale}}%
    \fi%
  \else%
    \setlength{\unitlength}{\svgwidth}%
  \fi%
  \global\let\svgwidth\undefined%
  \global\let\svgscale\undefined%
  \makeatother%
  \begin{picture}(1,0.32836877)%
    \put(0,0){\includegraphics[width=\unitlength,page=1]{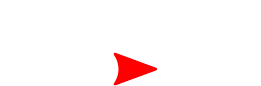}}%
    \put(0.50048819,0.22498336){\color[rgb]{0,0,0}\makebox(0,0)[b]{\smash{$k\,,ba$}}}%
    \put(0,0){\includegraphics[width=\unitlength,page=2]{Fermion_Propagator_Left.pdf}}%
  \end{picture}%
\endgroup%
}
\newcommand{\phononpropright}{%% Creator: Inkscape inkscape 0.91+devel+osxmenu, www.inkscape.org
\begingroup%
  \makeatletter%
  \providecommand\color[2][]{%
    \errmessage{(Inkscape) Color is used for the text in Inkscape, but the package 'color.sty' is not loaded}%
    \renewcommand\color[2][]{}%
  }%
  \providecommand\transparent[1]{%
    \errmessage{(Inkscape) Transparency is used (non-zero) for the text in Inkscape, but the package 'transparent.sty' is not loaded}%
    \renewcommand\transparent[1]{}%
  }%
  \providecommand\rotatebox[2]{#2}%
  \ifx\svgwidth\undefined%
    \setlength{\unitlength}{75.60131415bp}%
    \ifx\svgscale\undefined%
      \relax%
    \else%
      \setlength{\unitlength}{\unitlength * \real{\svgscale}}%
    \fi%
  \else%
    \setlength{\unitlength}{\svgwidth}%
  \fi%
  \global\let\svgwidth\undefined%
  \global\let\svgscale\undefined%
  \makeatother%
  \begin{picture}(1,0.33154477)%
    \put(0.49415118,0.22898166){\color[rgb]{0,0,0}\makebox(0,0)[b]{\smash{$q\,,\beta\alpha$}}}%
    \put(0,0){\includegraphics[width=\unitlength,page=1]{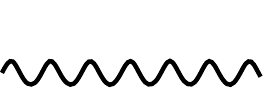}}%
  \end{picture}%
\endgroup%
}
\newcommand{\phononpropleft}{%% Creator: Inkscape inkscape 0.91+devel+osxmenu, www.inkscape.org
\begingroup%
  \makeatletter%
  \providecommand\color[2][]{%
    \errmessage{(Inkscape) Color is used for the text in Inkscape, but the package 'color.sty' is not loaded}%
    \renewcommand\color[2][]{}%
  }%
  \providecommand\transparent[1]{%
    \errmessage{(Inkscape) Transparency is used (non-zero) for the text in Inkscape, but the package 'transparent.sty' is not loaded}%
    \renewcommand\transparent[1]{}%
  }%
  \providecommand\rotatebox[2]{#2}%
  \ifx\svgwidth\undefined%
    \setlength{\unitlength}{75.60131415bp}%
    \ifx\svgscale\undefined%
      \relax%
    \else%
      \setlength{\unitlength}{\unitlength * \real{\svgscale}}%
    \fi%
  \else%
    \setlength{\unitlength}{\svgwidth}%
  \fi%
  \global\let\svgwidth\undefined%
  \global\let\svgscale\undefined%
  \makeatother%
  \begin{picture}(1,0.33154477)%
    \put(0.49415118,0.22898166){\color[rgb]{0,0,0}\makebox(0,0)[b]{\smash{$q\,,\beta\alpha$}}}%
    \put(0,0){\includegraphics[width=\unitlength,page=1]{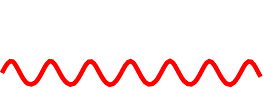}}%
  \end{picture}%
\endgroup%
}
\newcommand{\vertex}{%% Creator: Inkscape inkscape 0.91+devel+osxmenu, www.inkscape.org
\begingroup%
  \makeatletter%
  \providecommand\color[2][]{%
    \errmessage{(Inkscape) Color is used for the text in Inkscape, but the package 'color.sty' is not loaded}%
    \renewcommand\color[2][]{}%
  }%
  \providecommand\transparent[1]{%
    \errmessage{(Inkscape) Transparency is used (non-zero) for the text in Inkscape, but the package 'transparent.sty' is not loaded}%
    \renewcommand\transparent[1]{}%
  }%
  \providecommand\rotatebox[2]{#2}%
  \ifx\svgwidth\undefined%
    \setlength{\unitlength}{113.67675552bp}%
    \ifx\svgscale\undefined%
      \relax%
    \else%
      \setlength{\unitlength}{\unitlength * \real{\svgscale}}%
    \fi%
  \else%
    \setlength{\unitlength}{\svgwidth}%
  \fi%
  \global\let\svgwidth\undefined%
  \global\let\svgscale\undefined%
  \makeatother%
  \begin{picture}(1,0.74631675)%
    \put(0.23607233,0.67810662){\color[rgb]{0,0,0}\makebox(0,0)[b]{\smash{$q'\,,\alpha$}}}%
    \put(0,0){\includegraphics[width=\unitlength,page=1]{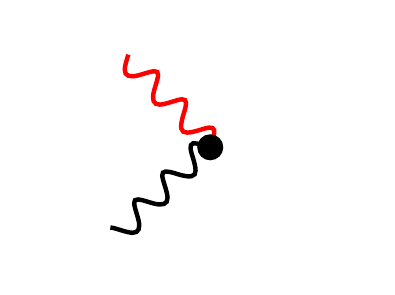}}%
    \put(0.86284955,0.67810662){\color[rgb]{0,0,0}\makebox(0,0)[b]{\smash{$k'\,,a$}}}%
    \put(0.23607233,0.01834114){\color[rgb]{0,0,0}\makebox(0,0)[b]{\smash{$q\,,\beta$}}}%
    \put(0.86284955,0.01834114){\color[rgb]{0,0,0}\makebox(0,0)[b]{\smash{$k\,,b$}}}%
    \put(0,0){\includegraphics[width=\unitlength,page=2]{FermionPhonon_Vertex.pdf}}%
  \end{picture}%
\endgroup%
}

\usepackage{setspace}

\makeatother

\usepackage{babel}
\begin{document}

\title{Nonlinear Bosonization and Refermionization in One Dimension with
the Keldysh Functional Integral}

\author{Filippo Bovo}
\maketitle

\lyxaddress{School of Physics and Astronomy, University of Birmingham, Edgbaston,
Birmingham, B15 2TT, UK.}
\begin{abstract}
We develop a self-contained approach to bosonization and refermionization
using the Keldysh functional integral. Starting from fermionic particles,
we bosonize the system and obtain a description in terms of the Tomonaga-Luttinger
liquid, with, in addition, an infinite series of interaction terms
arising from the curvature of the fermionic particle spectrum. We
explicitly calculate the leading interaction term and check its consistency
with a different approach based on the Matsubara framework, within
which we calculate the second leading interaction term, as well. Moreover,
we bosonize weakly and strongly interacting bosonic particles, and,
finally, refermionize interacting phonons into non-interacting fermionic
quasiparticles. The work culminates in a map between bosonic and fermionic
particles and effective bosonic and fermionic excitations, representing
phonons and fermionic quasiparticles.
\end{abstract}
\newpage{}

\section{Introduction}

Quantum systems in one dimension have been studied intensively with
regard to their static or equilibrium properties. Equilibrium properties
are associated with low energies, and one of the most important results
for quantum systems in one dimension is that their low-energy theory
can be expressed in terms of non-interacting phonons, even though
the particles constituting the system are interacting. This result
is known as bosonization \cite{Tomonaga1950,Luttinger1963,MattisLieb1965,Haldane1981a,Haldane1981b,Giamarchi},
and the phenomenology behind its success relies on this fact: particles
moving in one dimension cannot avoid each other as would be possible
in higher dimensions. As a consequence, any amount of interaction
between particles leads to a behaviour where each particle pushes
the next one in line in a sequence that leads to a density wave, that
is, a phonon.

Despite the fact that non-interacting phonons describe particularly
well one-dimensional systems in equilibrium, they are not sufficient
for a proper description of dynamical or non-equilibrium properties,
where higher energies become important. At zero temperatures, to capture
the dynamics it is sufficient to couple a single mobile impurity,
representing higher energies, to phonons, representing low energies
\cite{Giamarchi,ImambekovGlazman2012}. Instead, finite temperatures
require a thermodynamical number of mobile impurities \cite{ABG2014}.

The passage from static to dynamic properties and from zero to finite
temperatures requires a robust theoretical framework in order to capture
the properties of these systems. At zero temperature, the real time
many-body quantum field theory is a successful tool both for static
and dynamic properties \cite{ImambekovGlazman2012}. At finite temperatures,
the imaginary time Matsubara formalism is suitable for static properties
\cite{Giamarchi} and the Keldysh formalism is the natural candidate
for dynamic properties \cite{Kamenev2011}. Because of the recent
interest in the dynamics of one-dimensional quantum systems at finite
temperatures \cite{ImambekovGlazman2012}, in this work we derive
a self-contained framework for these systems within the Keldysh formalism\footnote{We will use Keldysh formalism and notation developed in Ref. \cite{Kamenev2011}.}.

We start with a phenomenological derivation of the hydrodynamic theory
of one-dimensional systems, describing low-energy excitations in terms
of phonons. After that, we turn to study fermionic particles, starting
from an overview of the spectrum of excitations of free fermions to
introduce important concepts to which we will refer in the rest of
the work. Using these concepts, we bosonize a system of interacting
fermions and obtain an equivalent description in terms of interacting
phonons. We proceed by evaluating the leading contribution, corresponding
to the Tomonaga-Luttinger liquid. Extending this result, we express
the system as the sum of the Tomonaga-Luttinger liquid, representing
non-interacting phonons, and an infinite series of terms, induced
by the curvature of the fermionic spectrum, describing interactions
between an arbitrary number of phonons. We explicitly calculate the
low-energy asymptotic expression of the leading interaction term,
representing the interaction between three phonons, and check the
consistency of this result in the Appendix using the Matsubara formalism.
In addition, in the Appendix, we derive the the low-energy asymptotic
expression of the next leading interaction term. Then we turn to a
system of bosonic particles with contact interactions and, in the
cases of weak and strong interactions, derive a description of their
low-energy excitations in terms of phonons. We conclude the work by
refermionizing the phononic systems obtained starting from fermions
or bosons to obtain a description of the low-energy excitations in
terms of free fermionic quasiparticles. Finally, we gather the results
into a map that translates between bosonic and fermionic particles
and effective bosonic and fermionic excitations, representing phonons
and fermionic quasiparticles.

\section{Hydrodynamics\label{chap:One-dimensional-systems}}

In one dimension, systems of bosonic and fermionic particles behave
differently than in higher dimensions, as particles constrained to
move on a line cannot avoid each other. As a consequence, fermions,
even in absence of interactions, cannot go past each other because
of the Pauli exclusion principle and, similarly, for bosons with local
repulsive interactions \cite{Giamarchi,Haldane1981b}. These systems
are dominated by collisions and, for times longer than the time between
two consecutive collisions, are in a hydrodynamic regime. Since longer
times translate into lower energies, at low enough energies we expect
these systems to be described by the hydrodynamic Hamiltonian of a
liquid \cite{Popov1991},
\[
\hat{H}=\int\mathrm{d}x\,\left[\frac{1}{2}m\hat{n}(x)\hat{v}(x)^{2}+e_{0}[\hat{n}]-\mu\hat{n}(x)\right]\,,
\]
where $\hat{n}(x)$ and $\hat{v}(x)$ are the density and velocity
operators of the liquid, $m$ and $e_{0}[\hat{n}]$ are the mass and
the ground state energy per unit length and $\mu$ is the chemical
potential. The first and second terms in the square brackets are respectively
the kinetic and internal energies of the liquid.

The ground state of the liquid corresponds to $\hat{v}=0$ and $\hat{n}=n$,
where $n$ is the constant homogeneous density determined by $e'_{0}[n]=\mu$
and the prime symbol, $'$, denotes derivative with respect to the
argument. The low-energy physics is associated with small variations
of velocity, $\hat{v}$, and density, $\hat{\rho}$, over the ground
state values, $\hat{v}=0$ and $\hat{n}=n$. Expressing the density
as $\hat{n}=n+\hat{\rho}$, the Hamiltonian becomes,
\[
\hat{H}=\int\mathrm{d}x\,\left[\frac{1}{2}m(n+\hat{\rho})\hat{v}^{2}+e_{0}(n+\hat{\rho})-\mu(n+\hat{\rho})\right]\,,
\]
and expanding over the small variations, $\hat{\rho}$ and $\hat{v}$,
we have,
\begin{equation}
\hat{H}\approx L\left[e_{0}(n)-\mu n\right]+\int\mathrm{d}x\,\left[\frac{1}{2}\left(mn\hat{v}^{2}+\frac{1}{\kappa n^{2}}\hat{\rho}^{2}\right)+\frac{m}{2}\hat{\rho}\hat{v}^{2}+\frac{\alpha}{6}\hat{\rho}^{3}\right]\,,\label{eq:Hydrodynamic_Hamiltonian}
\end{equation}
where $\kappa=1/n^{2}e''_{0}[n]$ is the compressibility of the liquid
\cite{PinesNozieres}, $\alpha=e'''_{0}[n]$ and we neglected higher
order terms. It is common to express the hydrodynamic Hamiltonian
in terms of the phase fields $\theta(x,t)$ and $\phi(x,t)$, related
to density and velocity as \cite{Haldane1981a,Haldane1981b,Giamarchi,ImambekovGlazman2012},\footnote{Here we follow the convention of Refs. \cite{Haldane1981a,Haldane1981b}.
The convention of Refs. \cite{Giamarchi,ImambekovGlazman2012} is
obtained by the substitutions $\phi\rightarrow\theta$ and $\theta\rightarrow-\phi$
in our equations.}
\begin{equation}
\begin{aligned}\hat{\rho}(x,t) & =\frac{1}{\pi}\partial_{x}\hat{\theta}(x,t)\,,\\
\hat{v}(x,t) & =\frac{1}{m}\partial_{x}\hat{\phi}(x,t)\,.
\end{aligned}
\label{eq:Density-Velocity_phases}
\end{equation}
Substituting Eqs. (\ref{eq:Density-Velocity_phases}) in Hamiltonian
(\ref{eq:Hydrodynamic_Hamiltonian}) we obtain,
\begin{equation}
\begin{aligned}\hat{H} & =L\left[e_{0}(n)-\mu n\right]\\
 & +\int\mathrm{d}x\,\left[\frac{c}{2\pi}\left(K(\partial_{x}\hat{\phi})^{2}+\frac{1}{K}(\partial_{x}\hat{\theta})^{2}\right)+\frac{1}{2\pi m}\partial_{x}\hat{\theta}(\partial_{x}\hat{\phi})^{2}+\frac{\alpha}{6\pi^{3}}(\partial_{x}\hat{\theta})^{3}\right]\,,
\end{aligned}
\label{eq:Luttinger-liquid_Hamiltonian_plus_Corrections}
\end{equation}
where we defined the speed of sound, $c$, and the Luttinger parameter,
$K$, as \cite{Giamarchi},
\begin{equation}
\begin{aligned}c= & \frac{1}{\sqrt{\kappa mn}}\,,\\
\frac{K}{\pi}= & \sqrt{\frac{\kappa n^{3}}{m}}\,.
\end{aligned}
\label{eq:Hydrodynamic_Parameters}
\end{equation}
The quadratic part of Eq. (\ref{eq:Luttinger-liquid_Hamiltonian_plus_Corrections}),
is the Tomonaga-Luttinger liquid Hamiltonian describing linear waves
in one dimension \cite{Giamarchi,Tomonaga1950,Luttinger1963,Haldane1981a,Haldane1981b}.
The cubic terms are the non-linear part of the Hamiltonian and are
useful to the dynamics of one-dimensional systems, as we will see
later. In this work, we will see how the low-energy physics of bosons
and fermions with local repulsive interactions in one dimension is
described by hydrodynamics Hamiltonian (\ref{eq:Luttinger-liquid_Hamiltonian_plus_Corrections})
and how to map it into an effective theory of free fermionic quasiparticles.

\section{Fermion hydrodynamics\label{sec:Fermion-hydrodynamics}}

\begin{figure}
\centering
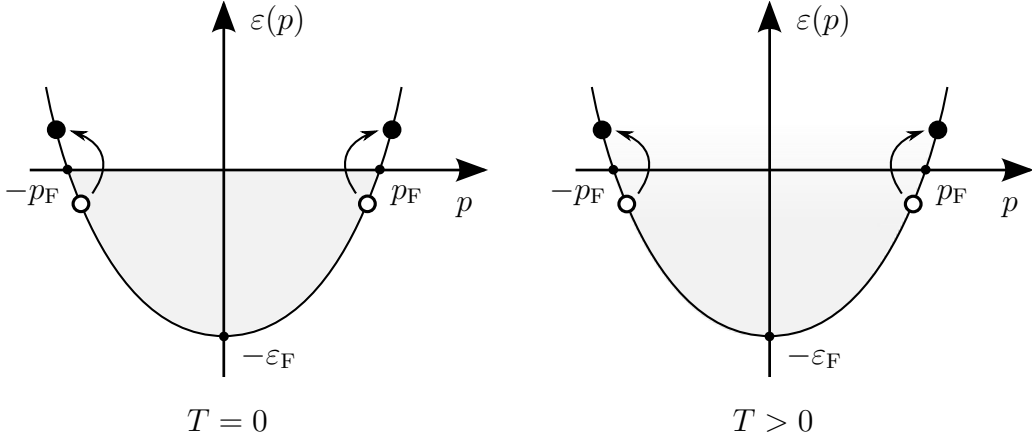

\caption[Occupation of states for free fermions at zero and finite temperatures.]{Occupation of states for free fermions at zero temperature, $T=0$,
and finite temperature, $T>0$. The Fermi sea is represented by the
grey regions and the Fermi surface is constituted by two separate
points, $p=\pm p_{\mathrm{F}}$. The temperature leads to a smearing
of the average occupation of states around the Fermi surface. The
circles and arrows represent particle-hole excitations. \label{fig:Spectrum_Free_Fermions}}
\end{figure}

In this section we derive the hydrodynamic theory for interacting
fermions. Since hydrodynamics is a theory of low-energy excitations,
it is instructive to study the low-energy excitations of free fermions
at zero and finite temperatures. At zero temperature, $T=0$, fermions
occupy all the states below the Fermi surface, as shown by the grey
region in Fig. \ref{fig:Spectrum_Free_Fermions}-left, forming the
Fermi sea.
 The Fermi surface in one dimension is constituted by two separate
points in momentum space, $p=\pm p_{\mathrm{F}}$, a central fact
for one dimensional hydrodynamics, as we will see soon. Fermions inside
the Fermi sea are frozen due to the Pauli exclusion principle. However,
they can jump across the Fermi points to the unfilled region above,
as shown by the arrows in Fig. \ref{fig:Spectrum_Free_Fermions}-left.
This process is referred to as particle-hole excitation. The hight
of the jump is the energy of the excitation and it becomes clear that
low-energy particle-hole excitations are fermions just below a Fermi
point jumping slightly above it. As a consequence, low-energy particle-hole
excitations must cross the Fermi points and remain close to them.
At temperatures much smaller than the Fermi energy, $T\ll\varepsilon_{\mathrm{F}}=\frac{p_{\mathrm{F}}^{2}}{2m}$,
where $m$ is the mass of fermions, the average occupation of the
states is smeared around the Fermi points, as shown by the grey region
in Fig. \ref{fig:Spectrum_Free_Fermions}-right. However, the above
argument still holds. Then, given that particle-hole excitations can
be arbitrarily close to the Fermi surface, they constitute the lowest
energy excitations of the system, suggesting that they could be good
candidates for the hydrodynamical description.

\begin{figure}
\centering
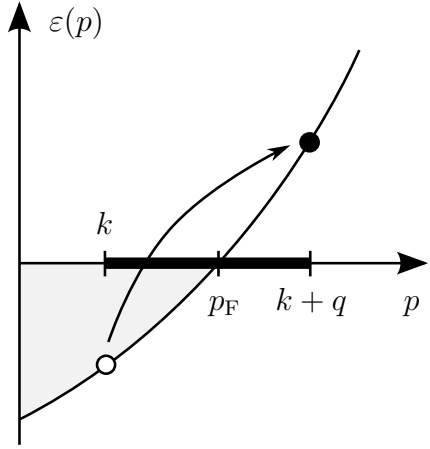

\caption[Particle-hole excitation close to the right Fermi point.]{Particle-hole excitation close to the right Fermi point, $p_{\mathrm{F}}$,
given by a fermion with momentum $k$ below the Fermi point excited
to a state with momentum $k+q$ above the Fermi point. The momentum
of the particle-hole excitation, $q$, is represented by the thick
line. The dashed line is the linearisation of the spectrum around
the Fermi point. \label{fig:Particle-Hole_exchitatioins}}
\end{figure}

\begin{figure}
\centering
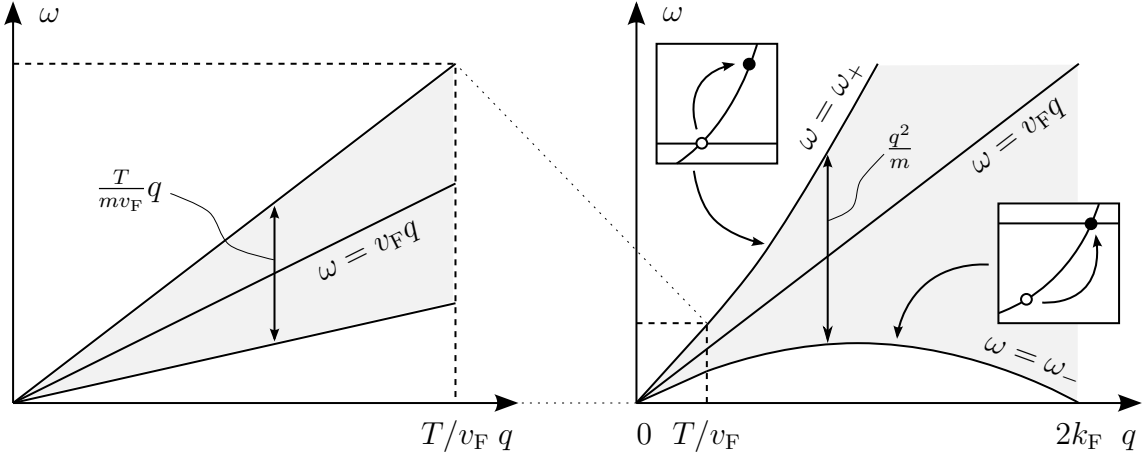

\caption[Dynamical structure factor of free fermions at a low temperatures.]{Dynamical structure factor, $S(q,\omega)$, of free fermions at a
low temperature, $T\ll\varepsilon_{\mathrm{F}}$. For $q\gg T/v_{\mathrm{F}}$,
$S(q,\omega)$ is given by the zero temperature result (\ref{eq:DSF_Free_Fermions_Zero_T})
and the grey colour represents the region where $S(q,\omega)$ is
non-zero. The grey region is centred around $\omega=v_{\mathrm{F}}q$
and is bounded above and below by $\omega=\omega_{+}(q)$ and $\omega=\omega_{-}(q)$,
the energies corresponding to the extremal particle-hole excitations
shown in the insets. For $q\ll T/v_{\mathrm{F}}$, $S(q,\omega)$
is given by the small temperature result (\ref{eq:DSF_Free_Fermions_Small_T})
and the grey colour represents qualitatively the width (\ref{eq:DSF_Free_Fermions_Small_T_width})
of bell-shaped distribution centred around $\omega=v_{\mathrm{F}}q$.
The small temperature result is magnified in the right picture.\label{fig:DSF}}
\end{figure}

To check this intuitive picture, we study the spectrum of particle-hole
excitations. For the moment, we use the operator formalism of second
quantisation, as it leads to a better insight. Although the particle-hole
excitation spectrum of free fermions is known \cite{ImambekovGlazman2012},
it is useful to review its calculation in order to build some intuition.
We start from the zero temperature case and later extend the results
to low temperatures. The spectrum of particle-hole excitations can
be studied using the dynamical structure factor, that gives the probability
per unit time to excite a density fluctuation of momentum $q$ and
energy $\omega$ by an external source. This quantity is accessible
through Bragg spectroscopy \cite{Ketterle1999} and was recently measured
for an array of one-dimensional Bose gases \cite{Clement2011}. The
dynamical structure factor is defined as the Fourier transform of
the density-density correlation \cite{PitaevskiiStringari,ImambekovGlazman2012},
\begin{equation}
S(q,\omega)=\int_{0}^{L}dx\int_{-\infty}^{\infty}dt\left\langle \hat{n}(x,t)\hat{n}(0,0)\right\rangle e^{-iqx+i\omega t}\,,\label{eq:DSF}
\end{equation}
where $\hat{n}(x,t)$ is the density operator, $L$ is the system
size and $\left\langle \ldots\right\rangle \equiv\langle0|\ldots|0\rangle$
denotes the expectation value over the ground state, $|0\rangle$.
Inserting the completeness relation $\sum_{j}|j\rangle\langle j|=1$,
where $j$ labels the energy eigenstates, between the densities in
Eq. (\ref{eq:DSF}), we obtain,

\[
S(q,\omega)=\int_{0}^{L}dx\int_{-\infty}^{\infty}dt\sum_{j}\bigl\langle0\bigr|\hat{n}(x,t)\bigl|j\bigr\rangle\bigl\langle j\bigr|\hat{n}(0,0)\bigl|0\bigr\rangle e^{-iqx+i\omega t}\,.
\]
Making the time dependence of the density explicit, $\hat{n}(x,t)=e^{i\hat{H}t}\hat{n}(x)e^{-i\hat{H}t}$,
we have,
\[
S(q,\omega)=\int_{0}^{L}dx\int_{-\infty}^{\infty}dt\sum_{j}e^{-i(E_{j}-E_{0})t}\bigl\langle0\bigr|\hat{n}(x)\bigl|j\bigr\rangle\bigl\langle j\bigr|\hat{n}(0)\bigl|0\bigr\rangle e^{-iqx+i\omega t}\,,
\]
where $E_{j}$ is the energy of the state $\left|j\right\rangle $
and $E_{0}$ is the energy of the ground state. Integrating over time
we find, 
\[
S(q,\omega)=2\pi\int_{0}^{L}dx\sum_{j}\bigl\langle0\bigr|\hat{n}(x)\bigl|j\bigr\rangle\bigl\langle j\bigr|\hat{n}(0)\bigl|0\bigr\rangle e^{-iqx}\delta(\omega-E_{j}+E_{0})\,.
\]
In terms of the Fourier transform of the fermionic creation and annihilation
operators of momentum $k$, $\hat{c}_{k}^{\dagger}$ and $\hat{c}_{k}$,
the density reads $\hat{n}(x)=\frac{1}{L}\sum_{q}\hat{n}_{q}e^{\mathrm{i}qx}=\frac{1}{L}\sum_{q,k}\hat{c}_{k-q}^{\dagger}\hat{c}_{k}e^{\mathrm{i}qx}$
and the dynamical structure factor becomes,
\begin{equation}
S(q,\omega)=\frac{2\pi}{L}\sum_{j}\sum_{q',k,k'}\bigl\langle0\bigr|\hat{c}_{k-q}^{\dagger}\hat{c}_{k}\bigl|j\bigr\rangle\bigl\langle j\bigr|\hat{c}_{k'-q'}^{\dagger}\hat{c}_{k'}\bigl|0\bigr\rangle\delta(\omega-E_{j}+E_{0})\,.\label{eq:DSF_Lehmann}
\end{equation}
Here, $\hat{c}_{k+q}^{\dagger}\hat{c}_{k}\left|0\right\rangle $ is
the ground state plus a particle excited from momentum $k$ to momentum
$k+q$, as pictured in Fig. \ref{fig:Particle-Hole_exchitatioins},
that is, a particle-hole excitation.
 Due to the orthogonality of these excited states, the only non-zero
contribution comes from $\left|j\right\rangle =\hat{c}_{k+q}^{\dagger}\hat{c}_{k}\bigl|0\bigr\rangle=\hat{c}_{k'+q'}^{\dagger}\hat{c}_{k'}\bigl|0\bigr\rangle$,
which also imposes $q=q'$ and $k=k'$. Then,
\[
\begin{aligned}S(q,\omega) & =\frac{2\pi}{L}\sum_{k}\left|\bigl\langle0\bigr|\hat{c}_{k}^{\dagger}\hat{c}_{k+q}\hat{c}_{k+q}^{\dagger}\hat{c}_{k}\bigl|0\bigr\rangle\right|^{2}\delta\left(\omega-\frac{q(q+2k)}{2m}\right)\\
 & \approx\frac{m}{q}\int\mathrm{d}k\,\left|\bigl\langle0\bigr|\hat{c}_{k}^{\dagger}\hat{c}_{k+q}\hat{c}_{k+q}^{\dagger}\hat{c}_{k}\bigl|0\bigr\rangle\right|^{2}\delta\left(k-\frac{m\omega}{q}+\frac{q}{2}\right)\,,
\end{aligned}
\]
where we approximated the sum by the integral and we used the fact
that the energy of the ground state plus a particle excited from momentum
$k$ to momentum $k-q$ is, 
\[
E_{j}=E_{0}+\frac{(k+q)^{2}}{2m}-\frac{k^{2}}{2m}=E_{0}+\frac{q(q+2k)}{2m}\,.
\]
In order for $c_{k+q}^{\dagger}c_{k}\left|0\right\rangle $ to be
different from zero, because of the Pauli exclusion principle, $c_{k}$
has to annihilate a particle inside the Fermi sea, that is, $|k|\leq k_{\mathrm{F}}$,
and $c_{k+q}^{\dagger}$ has to create a particle outside the Fermi
sea, that is, $|k+q|>k_{\mathrm{F}}$. In other words, in the case
of the right Fermi point, $p_{\mathrm{F}}$, the momentum $q$ of
the particle-hole excitation is represented by the thick line in Fig.
\ref{fig:Particle-Hole_exchitatioins} that can move left or right
but has to be pinned to the Fermi point. Considering only positive
$q$, for $q\leq2k_{\mathrm{F}}$ the restriction on $k$ becomes
$k_{\mathrm{F}}-q<k\leq k_{\mathrm{F}}$ and the result is symmetric
for negative $q$. Finally, integrating over $k$ we obtain,
\begin{equation}
\begin{aligned}S(q,\omega) & =\frac{m}{q}\theta\left(\omega-\omega_{-}(q)\right)\theta\left(\omega_{+}(q)-\omega\right)\,,\\
\omega_{\pm}(q) & =v_{\mathrm{F}}q\pm\frac{q^{2}}{2m}\,,
\end{aligned}
\label{eq:DSF_Free_Fermions_Zero_T}
\end{equation}
where $v_{\mathrm{F}}=p_{\mathrm{F}}/m$ is the Fermi velocity. The
dynamical structure factor at zero temperature as a function of $\omega$
is a box centred around $v_{\mathrm{F}}q$, of height $m/q$ and width,
\begin{equation}
\delta\omega(q)=\omega_{+}(q)-\omega_{-}(q)=\frac{q^{2}}{m}\,.\label{eq:DSF_width_zero_T}
\end{equation}
Particle-hole excitations with momentum $q$ and energy $\omega$
are present in the system where the dynamical structure factor is
non-zero, that is, the grey region in Fig. \ref{fig:DSF}-right (ignore
$q\ll T/v_{\mathrm{F}}$ for the moment).
 For each momentum, the dynamical structure factor is bound in energy
above and below. The upper bound, $\omega=\omega_{+}(q)$, corresponds
to a particle picked from the Fermi surface, $k_{\mathrm{F}}$, and
moved to $k_{\mathrm{F}}+q$ and the lower bound, $\omega=\omega_{-}(q)$,
corresponds to a particle picked from $k_{\mathrm{F}}-q$ and moved
to the Fermi surface, $k_{\mathrm{F}}$, as shown in the insets of
Fig. \ref{fig:DSF}-right. Higher and lower bounds correspond to the
thick line pinned to the Fermi surface in Fig. \ref{fig:Particle-Hole_exchitatioins}
moved completely to the right or left. Moving the thick line from
right to left creates all the intermediate particle-hole excitations
between higher and lower bounds, the grey region in Fig. \ref{fig:DSF}.
The peculiarity of one-dimensional systems is the absence of low-energy
excitations away from $q=0$ and $q=2k_{\mathrm{F}}$. This absence
is related to the presence of the lower bound, which in turn is related
to the Fermi surface in one dimension consisting of just two separate
points. In higher dimensions, Fermi surfaces are continuous, for example,
a circle in two dimensions, allowing one to play with angles to create
excitations with arbitrary low energy for any momentum \cite{Giamarchi}.

Now, we have a look at how the result is modified by the temperature.
At temperatures smaller than the Fermi energy, $T\ll\varepsilon_{\mathrm{F}}$,
the average over the ground state in Eq. (\ref{eq:DSF}) is replaced
by the thermal average $\sum_{i}e^{-E_{i}/T}\bigl\langle i\bigr|\ldots\bigl|i\bigr\rangle$,
and for $q\ll T/v_{\mathrm{F}}$ the dynamical structure factor becomes,\footnote{Note that to satisfy the $f$-sum rule, $\int\mathrm{d}\omega\,\omega S(q,\omega)\propto q^{2}/2m,$
(see e.g., Ref. \cite{PitaevskiiStringari}), one needs to consider
the full expression, the first line of Eq. (\ref{eq:DSF_Free_Fermions_Small_T}).
This is because the full expression correctly accounts for the detailed
balance at negative $\omega$.}
\begin{equation}
\begin{aligned}S(q,\omega) & =\frac{m}{q}n_{\mathrm{F}}\left(v_{\mathrm{F}}\frac{2m(\omega-v_{\mathrm{F}}q)-q^{2}}{2q}\right)n_{\mathrm{F}}\left(v_{\mathrm{F}}\frac{-2m(\omega-v_{\mathrm{F}}q)-q^{2}}{2q}\right)\\
 & \approx\frac{m}{4q}\frac{1}{\cosh^{2}\left(\frac{mv_{\mathrm{F}}}{2Tq}(\omega-v_{\mathrm{F}}q)\right)}\,.
\end{aligned}
\label{eq:DSF_Free_Fermions_Small_T}
\end{equation}
The procedure to obtain this result is similar to that at zero temperature
just seen and has been omitted. As a function of $\omega$, the dynamical
structure factor has the shape of a bell centred around $\omega=v_{\mathrm{F}}q$,
of height $m/4q$ and width linear in $q$,
\begin{equation}
\delta\omega_{T}(q)\sim\frac{T}{mv_{\mathrm{F}}}q\label{eq:DSF_Free_Fermions_Small_T_width}
\end{equation}
The width at finite temperatures is equal to the one at zero temperature,
Eq. (\ref{eq:DSF_width_zero_T}), with one power of $q$ replaced
by the thermal momentum, $T/v_{\mathrm{F}}$. The width of the dynamical
structure factor is represented by the grey region in Fig. \ref{fig:DSF}-left.
The thermal momentum scale, $T/v_{\mathrm{F}}$, separates a thermal
region for $q\ll T/v_{\mathrm{F}}$ from a quantum region for $q\gg T/v_{\mathrm{F}}$
as shown in \ref{fig:DSF}-left. The important observation is that
both at zero and finite temperature the widths $\delta\omega(q)$
and $\delta\omega_{T}(q)$ are small compared to the mean value $v_{\mathrm{F}}q$
in the small momentum and small energy limits, $q\ll k_{\mathrm{F}}$,
$\omega\ll\varepsilon_{\mathrm{F}}$ and $T\ll\varepsilon_{\mathrm{F}}$.
Referring to the zero temperature dynamical structure factor in Fig.
\ref{fig:DSF}, it means that the excitations are centred around the
line $\omega=v_{\mathrm{F}}q$. In turn, this means that in this limit
excitations have approximately fixed velocity $v_{\mathrm{F}}$, like
the speed of sound, $c$, of hydrodynamic wave excitation in Eq. (\ref{eq:Hydrodynamic_Parameters}).
Therefore, we have and intuitive link between low-energy fermions
and hydrodynamics. In the next two subsections we will see how the
physics of low-energy excitations around the Fermi points leads to
the hydrodynamic theory with speed of sound $c=v_{\mathrm{F}}$ for
free fermions and a modified speed of sound in presence of interactions.

\subsection{Bosonization}

We are ready to derive the hydrodynamic theory for interacting fermions,
a procedure known as bosonization \cite{Giamarchi,ImambekovGlazman2012}.
Bosonization of interacting fermions was first derived by Haldane
using the operator formalism of second quantisation \cite{Haldane1981a,Haldane1981b}.
Complementary to the operator approach, a functional integral approach
was suggested in Ref. \cite{Fogedby1976}, elaborated in Ref. \cite{LeeChen1988}
and presented in a clear picture in Refs. \cite{Yurkevich2002,GrishinYurkevichLerner2004,LernerYurkevich2005}.
This approach is known as functional bosonization and relies on the
Hubbard-Stratonovich transformations. All these approaches are based
on equilibrium physics and an extension to non-equilibrium was obtained
in Ref. \cite{MirlinEtAl2010} using the Keldysh technique. An appealing
aspect of bosonization is that the initial theory of interacting fermions
is reformulated as a non-interacting theory that is completely solvable.
However, the non-interacting theory is a low-energy approximation
to which refinements can be added. One of such refinement is the inclusion
of non-linear corrections. A non-linear correction was derived in
Ref. \cite{Haldane1981a} using the operator formalism and the assumption
that the excitations of the system are created above the ground state.
Here we will derive the functional bosonization using the Keldysh
technique and, in the spirit of Ref. \cite{kopietz1997}, a double
Hubbard-Stratonovich transformation approach, that results in a procedure
a bit different from those mentioned above. The advantage of this
procedure is that we are able to derive non-linear semiclassical corrections
within the functional integral formulation. The result is an infinite
series of terms where higher non-linear terms are less important.
We find that that the most important non-linear correction coincides
with the one calculated in Ref. \cite{Haldane1981a}. This correction
is derived using the Keldysh formalism. However, due to the complication
of the Keldysh indices and not to interrupt the flow of the section,
we derive rigorously more terms using the Matsubara formalism in Appendix
\ref{sec:N-loop-reduction-formula}.

Low-energy physics being restricted around the two Fermi points for
free fermions is at the base of bosonization. The addition of weak
local interactions between fermions smears the average occupation
around the Fermi points, in addition to the smearing due to the temperature.
However, the Fermi points, $\pm p_{\mathrm{F}}$, are not modified
by the presence of interactions as a consequence of the Luttinger's
theorem \cite{LuttingerWard1960,Luttinger1960,Tsvelik2003}. The Hamiltonian
of a system of interacting fermionic particles is,
\begin{equation}
\hat{H}=\int\mathrm{d}x\,\hat{\psi}^{\dagger}(x)\left[-\frac{\partial_{x}^{2}}{2m}-\varepsilon_{\mathrm{F}}\right]\hat{\psi}(x)+\frac{1}{2}\int\mathrm{d}x\mathrm{d}x'\,:\hat{\psi}^{\dagger}(x)\hat{\psi}(x)V(x-x')\hat{\psi}^{\dagger}(x')\hat{\psi}(x'):\,,\label{eq:Hamiltonian_Fermions_Interactions}
\end{equation}
where $\hat{\psi}(x)$ is the fermionic field operator, $m$ the mass
of fermions, $V(x-x')$ a local density-density interaction and the
colons denotes normal order of operators. In the functional integral
representation, the partition function corresponding to Hamiltonian
(\ref{eq:Hamiltonian_Fermions_Interactions}) is,
\begin{equation}
Z=\int\mathcal{D}\bigl[\bar{\psi},\psi\bigr]e^{\mathrm{i}\int_{\rightleftarrows}\mathrm{d}t\left[\int\mathrm{d}x\,\bar{\psi}(x,t)\left(\mathrm{i}\partial_{t}+\frac{\partial_{x}^{2}}{2m}+\varepsilon_{\mathrm{F}}\right)\psi(x,t)-\frac{1}{2}\int\mathrm{d}x\mathrm{d}x'\,\bar{\psi}(x,t)\psi(x,t)V(x-x')\bar{\psi}(x',t)\psi(x',t)\right]}\,,\label{eq:Partition_Function_Interacting_Fermions}
\end{equation}
where $\psi$ and $\bar{\psi}$ are the Grassmann fields corresponding
to $\hat{\psi}$ and $\hat{\psi}^{\dagger}$, $\rightleftarrows$
denotes the closed time contour\footnote{$\rightleftarrows$ corresponds to $\mathcal{C}$ in Ref. \cite{Kamenev2011}.},
and the thermal distribution in the infinite past has an average occupation,
\begin{equation}
n(p)=\frac{1}{e^{\beta(\varepsilon(p)-\varepsilon_{\mathrm{F}})}+1}\,,\label{eq:Occupation_factor_free_fermions}
\end{equation}
where $\varepsilon(p)=p^{2}/2m$ is the free fermionic spectrum. Now,
we use the fact that, at low energies, the physics is constrained
around the Fermi points in order to simplify the interaction term
in partition function (\ref{eq:Partition_Function_Interacting_Fermions}),
that in momentum space reads,
\begin{equation}
-\frac{1}{2}\int\mathrm{d}k\mathrm{d}k'\frac{\mathrm{d}q}{2\pi}\,V(q)\bar{\psi}(k-q,t)\psi(k,t)\bar{\psi}(k'+q,t)\psi(k',t)\,.\label{eq:Density-density_interaction_momentum_space}
\end{equation}
Here, quantities in real space and momentum space are related by,
\[
\begin{aligned}V(q) & =\int\mathrm{d}x\,V(x)e^{-\mathrm{i}qx}\,,\\
\psi(k,t) & =\int\mathrm{d}x\,\psi(x,t)e^{-\mathrm{i}kx}\,,
\end{aligned}
\]
where we considered a system of infinite length. The density-density
interaction term (\ref{eq:Density-density_interaction_momentum_space})
is represented by two particle-hole excitations: two particles with
momentum $k$ and $k'$ are annihilated and two with momentum $k-q$
and $k'+q$ are created.
\begin{figure}
\centering
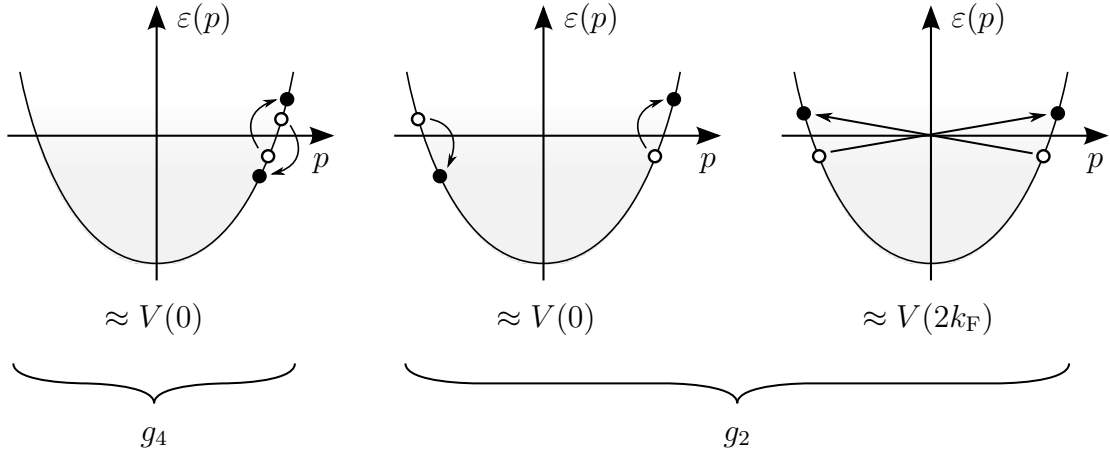

\caption[Relevant processes to the interaction term.]{Relevant processes to the interaction term (\ref{eq:Density-density_interaction_momentum_space}).
\label{fig:g-ology}}
\end{figure}
 At small energies, this leads to the three processes depicted in
Fig. \ref{fig:g-ology}. All the three processes have small total
momentum and energy; however, while single particle-hole excitations
in the first two processes have small momenta, $q\approx0$, the ones
in the third process have large momenta, $q\approx\pm2k_{\mathrm{F}}$,
as particles move to opposite Fermi points. It is then convenient
to split the fermion fields in the relevant parts, close to the Fermi
points, and irrelevant ones, away from them,
\begin{equation}
\begin{aligned}\psi(x,t) & =\int\frac{\mathrm{d}k}{2\pi}\,\psi(k,t)e^{\mathrm{i}kx}\\
 & =e^{\mathrm{i}k_{\mathrm{F}}x}\int_{-q_{0}/2}^{q_{0}/2}\frac{\mathrm{d}k}{2\pi}\,\psi(k_{\mathrm{F}}+k,t)e^{\mathrm{i}kx}+e^{-\mathrm{i}k_{\mathrm{F}}x}\int_{-q_{0}/2}^{q_{0}/2}\frac{\mathrm{d}k}{2\pi}\,\psi(-k_{\mathrm{F}}+k,t)e^{\mathrm{i}kx}+\ldots\\
 & =e^{\mathrm{i}k_{\mathrm{F}}x}\psi_{+}(x,t)+e^{-\mathrm{i}k_{\mathrm{F}}x}\psi_{-}(x,t)+\ldots\,,
\end{aligned}
\label{eq:Left-Right_field_Decomposition}
\end{equation}
where we introduced the momentum cut-off $q_{0}$ around the Fermi
points to delimit the relevant from the irrelevant parts, represented
by the dots. Since $\psi_{+}$ and $\psi_{-}$ are only constituted
by positive or negative momenta, they move respectively to the right
and to the left; for this reason we call them right and left movers.
Substituting decomposition (\ref{eq:Left-Right_field_Decomposition})
into partition function (\ref{eq:Partition_Function_Interacting_Fermions})
we have,
\begin{equation}
\begin{aligned}Z[u] & =\int\mathcal{D}\bigl[\bar{\psi}_{\pm},\psi_{\pm}\bigr]e^{\mathrm{i}\int_{\rightleftarrows}\mathrm{d}r\,\bar{\psi}_{\eta}(r)\left(G_{\eta}^{-1}(r)-u_{\eta}(r)\right)\psi_{\eta}(r)}\\
 & \hphantom{\int\mathcal{D}\bigl[\bar{\psi}_{\pm},\psi_{\pm}\bigr]}\times e^{-\frac{\mathrm{i}}{2}\int_{\rightleftarrows}\mathrm{d}r\,\bar{\psi}_{\eta}(r)\psi_{\eta}(r)g_{\eta\eta'}\bar{\psi}_{\eta'}(r)\psi_{\eta'}(r)}\,,
\end{aligned}
\label{eq:Left-Right_movers_partition_function}
\end{equation}
where $\eta=\pm1$, $r=(x,t)$, $\mathrm{d}r=\mathrm{d}x\mathrm{d}t$
and we defined the bare Green's functions,
\begin{equation}
G_{\eta}^{-1}(r)=\mathrm{i}\partial_{t}+\eta\mathrm{i}v_{\mathrm{F}}\partial_{x}+\frac{\partial_{x}^{2}}{2m}\,,\label{eq:Right-Left_movers_inverse_Green_Function}
\end{equation}
and the interaction matrix,
\[
\begin{aligned}g_{\eta\eta'} & =\left(\begin{array}{cc}
g_{4} & g_{2}\\
g_{2} & g_{4}
\end{array}\right)\,,\\
g_{4} & \approx V(0)\,,\\
g_{2} & \approx V(0)-V(2k_{F})\,.
\end{aligned}
\]
We also integrated over the fields far away from the Fermi points,
represented by the dots in decomposition (\ref{eq:Left-Right_field_Decomposition}),
as they contribute only through the quadratic free propagator to the
normalisation of the partition function. We added external source
fields, $u_{+}(r)$ and $u_{-}(r)$, coupled to the densities of right
and left movers, $\bar{\psi}_{+}(r)\psi_{+}(r)$ and $\bar{\psi}_{-}(r)\psi_{-}(r)$,
to keep track of what these quantities become after bosonization.
Now, we decouple the interaction term using a Hubbard-Stratonovich
transformation,
\[
Z[u]=\int\mathcal{D}\varrho_{\pm}e^{\frac{\mathrm{i}}{2}\int_{\rightleftarrows}\mathrm{d}r\,\varrho_{\eta}g_{\eta\eta'}^{-1}\varrho_{\eta'}}Z_{+}[\varrho_{+}+u_{+}]Z_{-}[\varrho_{-}+u_{-}]\,,
\]
where,
\[
Z_{\eta}[\varrho]=\int\mathcal{D}\bigl[\bar{\psi}_{\eta},\psi_{\eta}\bigr]e^{\mathrm{i}\int_{\rightleftarrows}\mathrm{d}r\,\bar{\psi}_{\eta}\left(G_{\eta}^{-1}-\varrho\right)\psi_{\eta}}\,,
\]
is the non-interacting part of the partition function. Here and in
the following we often omit the $r$ dependence of the fields. We
make the shift $\varrho\rightarrow\varrho-u$ and the partition function
becomes,
\begin{equation}
Z[u]=\int\mathcal{D}\varrho_{\pm}e^{\frac{\mathrm{i}}{2}\int_{\rightleftarrows}\mathrm{d}r\,\left(\varrho_{\eta}-u_{\eta}\right)g_{\eta\eta'}^{-1}\left(\varrho_{\eta'}-u_{\eta'}\right)}Z_{+}[\varrho_{+}]Z_{-}[\varrho_{-}]\,.\label{eq:Hubbard_Stratonovich_Density}
\end{equation}
The presence of the inverse interaction term is problematic when the
interaction is zero, in the case of free fermions. To avoid this,
in the spirit of Ref. \cite{kopietz1997}, we make a second Hubbard-Stratonovich
transformation,
\begin{equation}
Z[u]=\int\mathcal{D}\chi_{\pm}e^{\mathrm{i}\int_{\rightleftarrows}\mathrm{d}r\,\left[-\frac{1}{2}\chi'_{\eta}\frac{1}{(2\pi)^{2}}g_{\eta\eta'}\chi'_{\eta'}-\frac{1}{2\pi}\chi'_{\eta}u_{\eta}\right]}\tilde{Z}_{+}[\chi_{+}]\tilde{Z}_{-}[\chi_{-}]\,,\label{eq:Hubbard_Stratonovich_Density-1}
\end{equation}
where $\chi_{\pm}$ are called right and left chiral fields, prime
denotes a position derivative, $\chi'=\partial_{x}\chi$, and $\tilde{Z}_{\eta}$
is the functional Fourier transform of $Z_{\eta}$,
\[
\tilde{Z}_{\eta}[\chi_{\eta}]=\int\mathcal{D}\varrho_{\eta}e^{\mathrm{i}\int_{\rightleftarrows}\mathrm{d}r\,\frac{1}{2\pi}\chi'_{\eta}\varrho_{\eta}}Z_{\eta}[\varrho_{\eta}]\,.
\]
The reason for having $\partial_{x}$ acting on $\chi$ is that, as
we will see in the next subsection, the final result will be local
in position space.\footnote{The derivative also contributes to the normalisation of the partition
function, that we include in the integration measure.} After a Keldysh rotation, the partition function reads,
\begin{equation}
\begin{aligned}Z[u] & =\int\mathcal{D}\chi_{\pm}e^{\mathrm{i}\int\mathrm{d}r\,\left[-\frac{1}{2}\chi_{\eta}^{\prime\alpha}\left(\frac{1}{2\pi^{2}}g_{\eta\eta'}\right)\chi_{\eta'\alpha}^{\prime}-\frac{1}{\pi}\chi_{\eta}^{\prime\alpha}u_{\eta\alpha}\right]}\tilde{Z}_{+}[\chi_{+}]\tilde{Z}_{-}[\chi_{-}]\,,\\
\tilde{Z}_{\eta}[\chi_{\eta}] & =\int\mathcal{D}\varrho_{\eta}e^{\mathrm{i}\int\mathrm{d}r\,\frac{1}{\pi}\chi_{\eta}^{\prime\alpha}\varrho_{\eta\alpha}}Z_{\eta}[\varrho_{\eta}]\,,\\
Z_{\eta}[\varrho_{\eta}] & =\int\mathcal{D}\bigl[\bar{\psi}_{\eta},\psi_{\eta}\bigr]e^{\mathrm{i}\int_{-\infty}^{\infty}\mathrm{d}r\,\bar{\psi}_{\eta}^{a}\left[\left[G_{\eta}^{-1}\right]^{ab}-\varrho_{\eta}^{ab}\right]\psi_{\eta}^{b}}\,,
\end{aligned}
\label{eq:Partition_Function_Keldysh_Fermions_Full}
\end{equation}
where $\left[G_{\eta}^{-1}\right]^{ab}(x,t)=\delta^{ab}G_{\eta}^{-1}(x,t)$
is the inverse Green's function and $\varrho_{\eta}^{ab}=\delta^{ab}\varrho_{\eta}^{\mathrm{cl}}+\sigma_{1}^{ab}\varrho_{\eta}^{\mathrm{q}}$.
For brevity, we omit the classical and quantum indices in the integration
measure and we introduce the covariant-like notation $\varrho_{\alpha}=\sigma_{1}^{\alpha\beta}\varrho^{\beta}$,
where $\sigma_{1}^{\alpha\beta}$ is the first Pauli matrix. We also
omitted the infinitesimally small Keldysh component, $\left[G_{\eta}^{-1}\right]^{12}$
\cite{Kamenev2011}. Here and in the following we include the normalisation
factor of the partition function in the integration measure and use
the fact that $Z=1$ in absence of quantum sources. We evaluate $Z_{\eta}[\varrho_{\eta}]$
by integrating over right and left movers,
\begin{equation}
\begin{aligned}Z_{\eta}[\varrho_{\eta}] & =\frac{\det(-\mathrm{i}G_{\eta}^{-1}+\mathrm{i}\varrho_{\eta})}{\det(-\mathrm{i}G_{\eta}^{-1})}\\
 & =e^{\mathrm{Tr}\log(-\mathrm{i}G_{\eta}^{-1}+\mathrm{i}\varrho_{\eta})-\mathrm{Tr}\log(-\mathrm{i}G_{\eta}^{-1})}\\
 & =e^{\mathrm{Tr}\log(1-G_{\eta}\circ\varrho_{\eta})}\,,
\end{aligned}
\label{eq:Partition_function_Keldysh_Right-Left_movers}
\end{equation}
where the compact notation means,
\[
\begin{aligned}\varrho_{\eta} & =\varrho_{\eta}^{ab}(r)\delta^{2}(r-r')\,,\\
G_{\eta}^{-1} & =\left[G_{\eta}^{-1}\right]^{ab}(r)\delta^{2}(r-r')\,,\\
G_{\eta} & =G_{\eta}^{ab}(r-r')\,.
\end{aligned}
\]
Here $G_{\eta}$ is the right or left movers Green's function and
$\circ$ denotes convolution with respect to position and time. Substituting
result (\ref{eq:Partition_function_Keldysh_Right-Left_movers}) into
the second line of Eq. (\ref{eq:Partition_Function_Keldysh_Fermions_Full}),
we have,
\begin{equation}
\tilde{Z}_{\eta}[\chi_{\eta}]=\int\mathcal{D}\varrho_{\eta}e^{\mathrm{i}\int\mathrm{d}r\,\frac{1}{\pi}\chi_{\eta}^{\prime\alpha}\varrho_{\eta\alpha}+\mathrm{Tr}\log(1-G_{\eta}\circ\varrho_{\eta})}\,.\label{eq:Partition_Function_chi_Fourier_Transform}
\end{equation}
The trace of the logarithm can be expanded in the usual way \cite{AltlandSimons2010Book},
leading to the n-particle vertices for the field $\varrho_{\eta}$,
\begin{figure}
\centering
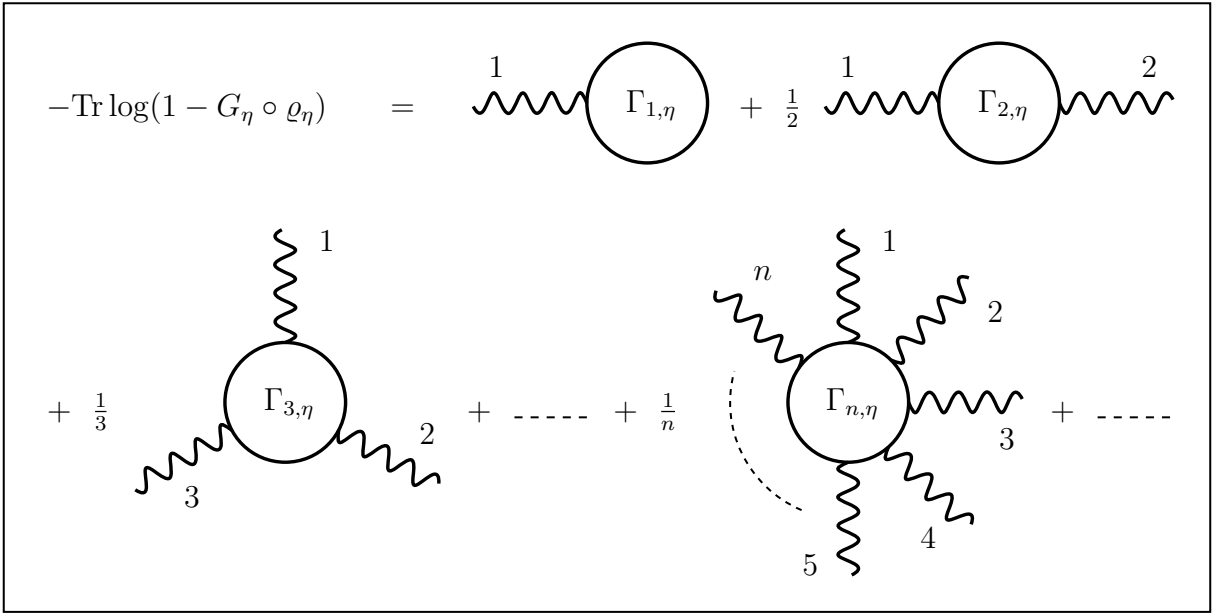

\caption[Diagrammatic representation of the n-loop series.]{Diagrammatic representation of series (\ref{eq:Bosonization_Series}).
The wavy lines represent the fields $\varrho_{\eta}$'s and the solid
lines the fermionic Green's functions, $G_{\eta}$'s. The fermionic
loops $\Gamma_{n,\eta}$ are given by Eq. (\ref{eq:n-vertex}). \label{fig:Vertex_Corrections}}
\end{figure}

\begin{equation}
\begin{aligned}-\mathrm{Tr}\log(1-G_{\eta}\circ\varrho_{\eta}) & =\mathrm{Tr}(G_{\eta}\circ\varrho_{\eta})+\frac{1}{2}\mathrm{Tr}(G_{\eta}\circ\varrho_{\eta}\circ G_{\eta}\circ\varrho_{\eta})+\ldots\\
 & =\sum_{n\geq1}\frac{1}{n}\int\mathrm{d}r_{1}\ldots\mathrm{d}r_{n}\,\Gamma_{n\eta}^{\alpha_{1}\ldots\alpha_{n}}(r_{1},\ldots,r_{n})\varrho_{\eta}^{\alpha_{1}}(r_{1})\cdots\varrho_{\eta}^{\alpha_{n}}(r_{n})\,.
\end{aligned}
\label{eq:Bosonization_Series}
\end{equation}
The interaction vertex is,
\begin{equation}
\Gamma_{n,\eta}^{\alpha_{1}\ldots\alpha_{n}}(r_{1},\ldots,r_{n})=\sigma_{\alpha_{1}}^{a_{1},a'_{1}}G_{\eta}^{a'_{1}a_{2}}(r_{1}-r_{2})\sigma_{\alpha_{2}}^{a_{2},a'_{2}}G_{\eta}^{a'_{2}a_{3}}(r_{2}-r_{3})\cdots\sigma_{\alpha_{n}}^{a_{n},a'_{n}}G_{\eta}^{a'_{n}a_{1}}(r_{n}-r_{1})\,,\label{eq:n-vertex}
\end{equation}
where $\sigma_{\alpha}^{a,a'}=\delta_{\alpha,\mathrm{cl}}\delta^{aa'}+\delta_{\alpha,\mathrm{q}}\sigma_{1}^{aa'}$.
Series (\ref{eq:Bosonization_Series}) is represented by means of
the Feynman diagrams in Fig. \ref{fig:Vertex_Corrections}. The wavy
lines represent the fields $\varrho_{\eta}$'s and the solid lines
the free fermion Green functions, $G_{\eta}$'s. The first term of
the series, known as tadpole diagram, is $\sim\int\mathrm{d}r\varrho_{\eta}^{\mathrm{q}}(r)$
and just contributes to the homogeneous density of the system \cite{ZinnJustin}.
We omit it by measuring the density from its constant homogeneous
value. The second term of the series, know as polarisation diagram,
contributes to the free propagator of $\varrho_{\eta}$ and is the
leading contribution in the bosonization procedure. The rest of the
terms, $n\geq3$, are the non-linear diagrams, generating interactions
between the fields $\varrho_{\eta}$'s. Although the series of diagrams
is infinite, in the limit of low energies non-linear corrections become
smaller compared to the polarisation diagram, as we will see in the
following. We proceed by considering the lowest approximation, that
is, the polarisation diagram.

\subsection{Linear spectrum}

\begin{figure}
\centering
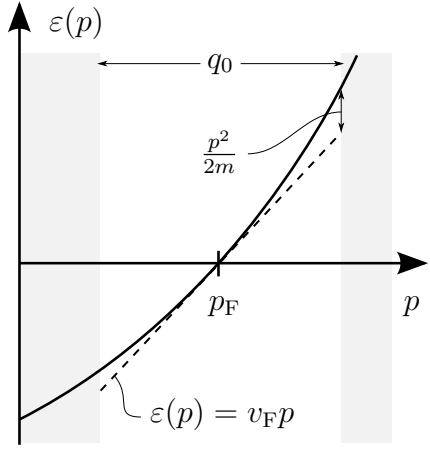

\caption[Linearisation of the fermionic spectrum around the right Fermi point.]{Linearisation of the spectrum, $\varepsilon(p)=v_{\mathrm{F}}p+p^{2}/2m$,
around the right Fermi point. The dashed line represents the linearised
spectrum, $\varepsilon(p)=v_{\mathrm{F}}p$, and $q_{0}$ is the cut-off
introduced in the decomposition of the fermionic field into left and
right movers, Eq. (\ref{eq:Left-Right_field_Decomposition}). \label{fig:Linearization_Spectrum}}
\end{figure}

The lower the energy, the lower the momentum and the smaller the relevant
region around the Fermi points. This means that we may choose a smaller
cut-off, $q_{0}$. The spectrum with the cut-off around the right
Fermi point is represented in Fig. \ref{fig:Linearization_Spectrum}.
The dashed line in the figure represents the linearised spectrum around
the right Fermi point and it is clear that the smaller the cut-off,
the smaller the contribution of the curvature. Then, in the limit
$q_{0}\ll mv_{\mathrm{F}}$, in first approximation we may neglect
the curvature. To do so, we consider the fermionic spectrum with momentum
centred around the right or left Fermi points, $\pm p_{\mathrm{F}}$,
\[
\varepsilon(p)=\frac{p^{2}}{2m}-\varepsilon_{\mathrm{F}}\longrightarrow\varepsilon_{\eta}(p)=\frac{(p-\eta p_{\mathrm{F}})^{2}}{2m}-\frac{p_{\mathrm{F}}^{2}}{2m}=\eta v_{\mathrm{F}}p+\frac{p^{2}}{2m}\,.
\]
The linear approximation amounts to neglecting the quadratic term,
retaining only the linear one,
\[
\varepsilon_{\eta}(p)\approx\eta v_{\mathrm{F}}p\,.
\]
In the case of linear spectrum, Dzyaloshinski and Larkin proved that
in equilibrium only the first two diagram in Fig. \ref{fig:Vertex_Corrections}
are non-zero, that is, $\Gamma_{\eta,n}=0$ for $n\geq3$ in Eqs.
(\ref{eq:Bosonization_Series}) and (\ref{eq:n-vertex}) \cite{DzialoshinskiiLarkin1974}.
This statement goes by the name of the Dzyaloshinski-Larkin theorem.\label{subsec:Dzyaloshinski-Larking theorem}
In the non-equilibrium case, an infinite series of non-linear terms
in the quantum field is present, leading to an expansion in noise
cumulants \cite{MirlinEtAl2010}. However, we neglect these terms
by considering small deviations from equilibrium. Taking the linear
spectrum approximation and using the Dzyaloshinski-Larkin theorem,
Eq. (\ref{eq:Partition_Function_chi_Fourier_Transform}) becomes,
\[
\begin{aligned}\tilde{Z}_{\eta}[\chi_{\eta}] & =\int\mathcal{D}\varrho_{\eta}e^{\mathrm{i}\int\mathrm{d}r\,\frac{1}{\pi}\chi_{\eta\alpha}^{\prime}\varrho_{\eta}^{\alpha}-\frac{\mathrm{i}}{2}\int\mathrm{d}r\mathrm{d}r'\,\varrho_{\eta}^{\alpha}(r)\Pi_{\eta}^{\alpha\beta}(r-r')\varrho_{\eta}^{\beta}(r')}\\
 & =\frac{1}{\sqrt{\det\mathrm{i}\Pi_{\eta}}}e^{\mathrm{i}\int\mathrm{d}r\mathrm{d}r'\,\frac{1}{2}\chi_{\eta\alpha}^{\prime}(r)\frac{1}{\pi^{2}}\left[\Pi_{\eta}^{-1}\right]^{\alpha\beta}(r-r')\chi_{\eta\beta}^{\prime}(r')}\,,
\end{aligned}
\]
where,
\begin{equation}
\mathrm{i}\Pi_{\eta}^{\alpha\beta}(r)=\Gamma_{2,\eta}^{\alpha\beta}(r,-r)=\left(\begin{array}{cc}
0 & \mathrm{i}\Pi_{\eta}^{\mathrm{A}}(r)\\
\mathrm{i}\Pi_{\eta}^{\mathrm{R}}(r) & \mathrm{i}\Pi_{\eta}^{\mathrm{K}}(r)
\end{array}\right)\,,\label{eq:Polarization_Keldysh}
\end{equation}
is the polarisation function of free fermions and,
\[
\begin{aligned}\mathrm{i}\Pi_{\eta}^{\mathrm{R}}(x,t) & =\left[\mathrm{i}\Pi_{\eta}^{\mathrm{A}}(-x,-t)\right]^{*}\\
 & =G_{\eta}^{\mathrm{R}}(x,t)G_{\eta}^{\mathrm{K}}(-x,-t)+G_{\eta}^{\mathrm{K}}(x,t)G_{\eta}^{\mathrm{A}}(-x,-t)\,,\\
\mathrm{i}\Pi_{\eta}^{\mathrm{K}}(x,t) & =G_{\eta}^{\mathrm{K}}(x,t)G_{\eta}^{\mathrm{K}}(-x,-t)+G_{\eta}^{\mathrm{R}}(x,t)G_{\eta}^{\mathrm{A}}(-x,-t)+G_{\eta}^{\mathrm{A}}(x,t)G_{\eta}^{\mathrm{R}}(-x,-t)\\
 & =G_{\eta}^{\mathrm{K}}(x,t)G_{\eta}^{\mathrm{K}}(-x,-t)-\left(G_{\eta}^{\mathrm{R}}(x,t)-G_{\eta}^{\mathrm{A}}(x,t)\right)\left(G_{\eta}^{\mathrm{R}}(x,t)-G_{\eta}^{\mathrm{A}}(x,t)\right)\,,
\end{aligned}
\]
where in the last line we used,
\begin{equation}
\begin{aligned} & G_{\eta}^{\mathrm{R}}(x,t)G_{\eta}^{\mathrm{R}}(-x,-t)=0\,,\\
 & G_{\eta}^{\mathrm{A}}(x,t)G_{\eta}^{\mathrm{A}}(-x,-t)=0\,,
\end{aligned}
\label{eq:Support_Null_Dimension}
\end{equation}
since their support is of null dimension \cite{Kamenev2011}. The
polarisation function is easier to calculate in the Fourier representation,
\[
\varrho_{\eta}^{\alpha}(x,t)=\int\frac{\mathrm{d}q}{2\pi}\frac{\mathrm{d}\omega}{2\pi}\varrho_{\eta}^{\alpha}(q,\omega)e^{\mathrm{i}qx-\mathrm{i}\omega t}\,,
\]
that gives,
\begin{equation}
\tilde{Z}_{\eta}[\chi_{\eta}]=\frac{1}{\sqrt{\det\mathrm{i}\Pi_{\eta}}}e^{\mathrm{i}\int\frac{\mathrm{d}\omega}{2\pi}\frac{\mathrm{d}q}{2\pi}\,\frac{1}{2}\bar{\chi}_{\eta\alpha}(q,\omega)\frac{q^{2}}{\pi^{2}}\left[\Pi_{\eta}^{-1}\right]^{\alpha\beta}(q,\omega)\chi_{\eta\beta}(q,\omega)}\,,\label{eq:Partition_Function_Polarization}
\end{equation}
where,
\[
\begin{aligned}\Pi_{\eta}^{\mathrm{R}}(q,\omega) & =\left[\Pi_{\eta}^{\mathrm{A}}(q,\omega)\right]^{*}\\
 & =-\mathrm{i}\int\frac{\mathrm{d}k}{2\pi}\frac{\mathrm{d}\varepsilon}{2\pi}\left[G_{\eta}^{\mathrm{R}}(k+q,\varepsilon+\omega)G_{\eta}^{\mathrm{K}}(k,\varepsilon)+G_{\eta}^{\mathrm{K}}(k+q,\varepsilon+\omega)G_{\eta}^{\mathrm{A}}(k,\varepsilon)\right]\\
 & =\mathrm{i}\int\frac{\mathrm{d}k}{2\pi}\left(F(\varepsilon_{\eta}(k))-F(\varepsilon_{\eta}(k+q)\right)\int\frac{\mathrm{d}\varepsilon}{2\pi}G_{\eta}^{\mathrm{R}}(k+q,\varepsilon+\omega)G_{\eta}^{\mathrm{A}}(k,\varepsilon)\,,\\
\Pi_{\eta}^{\mathrm{K}}(q,\omega) & =-\mathrm{i}\int\frac{\mathrm{d}k}{2\pi}\frac{\mathrm{d}\varepsilon}{2\pi}\left[G_{\eta}^{\mathrm{K}}(k+q,\varepsilon+\omega)G_{\eta}^{\mathrm{K}}(k,\varepsilon)\right.\\
 & \quad\left.-\left(G_{\eta}^{\mathrm{R}}(k+q,\varepsilon+\omega)-G_{\eta}^{\mathrm{A}}(k+q,\varepsilon+\omega)\right)\left(G_{\eta}^{\mathrm{R}}(k,\varepsilon)-G_{\eta}^{\mathrm{A}}(k,\varepsilon)\right)\right]\\
 & =-\mathrm{i}\int\frac{\mathrm{d}k}{2\pi}\left(F(\varepsilon_{\eta}(k+q))F(\varepsilon_{\eta}(k))-1\right)\\
 & \quad\times\frac{\mathrm{d}\varepsilon}{2\pi}\left(G_{\eta}^{\mathrm{R}}(k+q,\varepsilon+\omega)-G_{\eta}^{\mathrm{A}}(k+q,\varepsilon+\omega)\right)\left(G_{\eta}^{\mathrm{R}}(k,\varepsilon)-G_{\eta}^{\mathrm{A}}(k,\varepsilon)\right)\,.
\end{aligned}
\]
Here we used the fluctuation-dissipation theorem to express the Keldysh
component in terms of retarded and advanced ones,
\[
\begin{aligned}G_{\eta}^{\mathrm{K}}(k,\varepsilon) & =F(\varepsilon)\left(G_{\eta}^{\mathrm{R}}(k,\varepsilon)-G_{\eta}^{\mathrm{A}}(k,\varepsilon)\right)\\
 & =F(\varepsilon_{\eta}(k))\left(G_{\eta}^{\mathrm{R}}(k,\varepsilon)-G_{\eta}^{\mathrm{A}}(k,\varepsilon)\right)\,.
\end{aligned}
\]
We also used the fact that Eqs. (\ref{eq:Support_Null_Dimension})
imply,
\[
\begin{aligned} & \int\frac{\mathrm{d}\varepsilon}{2\pi}G_{\eta}^{\mathrm{R}}(k+q,\varepsilon+\omega)G_{\eta}^{\mathrm{R}}(k,\varepsilon)=0\,,\\
 & \int\frac{\mathrm{d}\varepsilon}{2\pi}G_{\eta}^{\mathrm{A}}(k+q,\varepsilon+\omega)G_{\eta}^{\mathrm{A}}(k,\varepsilon)=0\,.
\end{aligned}
\]
Using the thermal equilibrium average occupation, $F(\varepsilon)=\tanh\left(\frac{\varepsilon}{2T}\right)$,
and integrating over $\varepsilon$, the retarded component of the
polarisation function becomes,
\[
\begin{aligned}\Pi_{\eta}^{\mathrm{R}}(q,\omega) & =\frac{1}{\omega-\eta cq+\mathrm{i}0}\int\frac{\mathrm{d}k}{2\pi}\left[\tanh\left(\eta\frac{v_{\mathrm{F}}k+v_{\mathrm{F}}q}{2T}\right)-\tanh\left(\eta\frac{v_{\mathrm{F}}k}{2T}\right)\right]\\
 & =\frac{\eta}{\pi}\frac{q}{\omega-\eta cq+\mathrm{i}0}\,,
\end{aligned}
\]
and the Keldysh component,
\[
\begin{aligned}\Pi_{\eta}^{\mathrm{K}}(q,\omega) & =-\mathrm{i}\delta(\omega-\eta v_{\mathrm{F}}q)\int\mathrm{d}k\left[1-\tanh\left(\eta\frac{v_{\mathrm{F}}k+v_{\mathrm{F}}q}{2T}\right)\tanh\left(\eta\frac{v_{\mathrm{F}}k}{2T}\right)\right]\\
 & =-\mathrm{i}\delta(\omega-\eta v_{\mathrm{F}}q)\coth\left(\frac{cq}{2T}\right)\int\mathrm{d}k\left[\tanh\left(\frac{v_{\mathrm{F}}k+v_{\mathrm{F}}q}{2T}\right)-\tanh\left(\frac{v_{\mathrm{F}}k}{2T}\right)\right]\\
 & =-2\mathrm{i}q\coth\left(\frac{v_{\mathrm{F}}q}{2T}\right)\delta(\omega-\eta v_{\mathrm{F}}q)\,,
\end{aligned}
\]
where in the second line we used the identity,
\begin{equation}
1-\tanh(a)\tanh(b)=\coth\left(a-b\right)\left[\tanh(a)-\tanh(b)\right]\,.\label{eq:MagicIdentityTanh}
\end{equation}
Summarising, retarded, advanced and Keldysh Green's functions are,
\[
\begin{aligned}\Pi_{\eta}^{\mathrm{R},\mathrm{A}}(q,\omega) & =\frac{\eta}{\pi}\frac{q}{\omega-\eta v_{\mathrm{F}}q\pm\mathrm{i}0}\,,\\
\Pi_{\eta}^{\mathrm{K}}(q,\omega) & =-2\mathrm{i}q\coth\left(\frac{v_{\mathrm{F}}q}{2T}\right)\delta(\omega-\eta v_{\mathrm{F}}q)\,,\\
 & =\coth\left(\frac{\omega}{2T}\right)\left(\Pi_{\eta}^{\mathrm{R}}(q,\omega)-\Pi_{\eta}^{\mathrm{A}}(q,\omega)\right)\,,
\end{aligned}
\]
where the last identity is a statement of the fluctuation-dissipation
theorem. Now that we know the explicit form of the polarisation function,
Eq. (\ref{eq:Polarization_Keldysh}), we can invert it and substitute
it in Eq. (\ref{eq:Partition_Function_Polarization}). The inverted
polarisation Green's function reads, 
\[
\left[\Pi_{\eta}^{-1}(q,\omega)\right]^{\alpha\beta}=\Pi_{\eta}^{-1}(q,\omega)\sigma_{1}^{\alpha\beta}=\eta\pi\frac{\omega-\eta v_{\mathrm{F}}q}{q}\sigma_{1}^{\alpha\beta}\,,
\]
where, as before, we omitted the infinitesimally small Keldysh component
\cite{Kamenev2011}. Substituting the inverse Green's function in
Eq. (\ref{eq:Partition_Function_Polarization}) we have,
\[
\tilde{Z}_{\eta}[\chi_{\eta}]=\frac{1}{\sqrt{\det\mathrm{i}\Pi_{\eta}}}e^{\mathrm{i}\int\frac{\mathrm{d}\omega}{2\pi}\frac{\mathrm{d}q}{2\pi}\,\frac{1}{2}\bar{\chi}_{\eta}^{\alpha}(q,\omega)D_{0,\eta}^{-1}(q,\omega)\chi_{\eta\alpha}(q,\omega)}\,,
\]
where we defined the propagator of the chiral fields without the contribution
$\sim g_{\eta\eta'}$ from the density-density interactions of fermions
(see first line of Eq. (\ref{eq:Partition_Function_Keldysh_Fermions_Full}))
as,
\[
D_{0,\eta}^{-1}(q,\omega)=\frac{q^{2}}{\pi^{2}}\Pi_{\eta}^{-1}(q,\omega)=\frac{\eta}{\pi}q(\omega-\eta v_{\mathrm{F}}q)\,.
\]
Fourier transforming back to position and time, the partition function
becomes,
\begin{equation}
Z_{\eta}[u]=\frac{1}{\sqrt{\det\mathrm{i}\Pi_{\eta}}}e^{\mathrm{i}\int\mathrm{d}r\,\frac{1}{2}\chi_{\eta}^{\alpha}(r)D_{0,\eta}^{-1}(r)\chi_{\eta\alpha}(r)}\,,\label{eq:Partition_Function_Chi_2}
\end{equation}
where,
\begin{equation}
D_{0,\eta}^{-1}(r)=\frac{\eta}{\pi}\partial_{x}(\partial_{t}+\eta v_{\mathrm{F}}\partial_{x})\,,\label{eq:Chi_Propagator_no_g}
\end{equation}
Finally, substituting Eqs. (\ref{eq:Partition_Function_Chi_2}) into
the first line in Eq. (\ref{eq:Partition_Function_Keldysh_Fermions_Full}),
we find the bosonized partition function of fermions with linear spectrum,
\begin{equation}
Z[u]=\int\mathcal{D}\chi_{\pm}e^{\mathrm{i}\int\mathrm{d}r\,\left[\frac{1}{2}\chi_{\eta}^{\alpha}D_{\eta\eta'}^{-1}\chi_{\eta'\alpha}-\frac{1}{\pi}\chi_{\eta}^{\prime\alpha}u_{\eta\alpha}\right]}\,,\label{eq:Partition_Function_chi}
\end{equation}
where the chiral fields propagator is,
\begin{equation}
D_{\eta\eta'}^{-1}(r)=\delta_{\eta\eta'}D_{0,\eta}^{-1}(r)+\frac{1}{\pi}\frac{g_{\eta\eta'}}{2\pi}\partial_{x}^{2}\,.\label{eq:Chi_Propagator}
\end{equation}
The first thing that we note is that the exponent of partition function
(\ref{eq:Partition_Function_chi}) is quadratic in $\chi_{\pm}$.
As hinted at the beginning of the section, this is a major advantage
of bosonization: we started with a system of interacting electrons
and we ended up with a quadratic low-energy theory that allows for
exact analytical results. Thanks to the external source $u_{\eta}$
we can relate the right and left chiral fields with right and left
movers by comparing Eqs. (\ref{eq:Left-Right_movers_partition_function})
and (\ref{eq:Partition_Function_chi}), to obtain the correspondence,\footnote{Note that in this definition we dropped the Keldysh indices by writing
the fields in the closed time contour, $\rightleftarrows$, in order
to meet the standard definitions found in literature \cite{Haldane1981a,Haldane1981b,Giamarchi,ImambekovGlazman2012}.
We also removed a factor 2 multiplying $\partial_{x}\chi_{\eta}^{\alpha}/2\pi$
that appears Keldysh formalism (See also Ref. \cite{Kamenev2011}).
We will refer again to this footnote, for example for the fields $\theta$
and $\phi$.\label{fn:KeldyshFactor2}}
\begin{equation}
\bar{\psi}_{\eta}\psi_{\eta}\longleftrightarrow\frac{1}{2\pi}\partial_{x}\chi_{\eta}\,.\label{eq:Movers_Chiral_relation}
\end{equation}
This means that $\frac{1}{2\pi}\partial_{x}\chi_{\pm}$ has the meaning
of density of right and left movers. Partition function (\ref{eq:Partition_Function_chi})
is not the usual form found in literature \cite{Giamarchi}. The usual
form corresponds to the expression in parenthesis of Eq. (\ref{eq:Luttinger-liquid_Hamiltonian_plus_Corrections})
and, to obtain it, we make the transformation,
\begin{equation}
\chi_{\pm}=\theta\pm\phi\,,\label{eq:Chiral2PhaseDensity}
\end{equation}
and find the Tomonaga-Luttinger liquid partition function,
\begin{equation}
Z[u]=\int\mathcal{D}\theta\mathcal{D}\phi e^{\mathrm{i}\int\mathrm{d}r\,\left[\left(\begin{array}{c}
\theta^{\alpha}\\
\phi^{\alpha}
\end{array}\right)^{\mathrm{T}}\left(\begin{array}{cc}
\frac{c}{\pi K}\partial_{x}^{2} & \frac{1}{\pi}\partial_{t}\partial_{x}\\
\frac{1}{\pi}\partial_{t}\partial_{x} & \frac{cK}{\pi}\partial_{x}^{2}
\end{array}\right)\left(\begin{array}{c}
\theta_{\alpha}\\
\phi_{\alpha}
\end{array}\right)-\left(\begin{array}{c}
\frac{2}{\pi}\partial_{x}\theta^{\alpha}\\
\frac{2}{\pi}\partial_{x}\phi^{\alpha}
\end{array}\right)\left(\begin{array}{c}
u_{\theta\alpha}\\
u_{\phi\alpha}
\end{array}\right)\right]}\label{eq:Fermion_Bosonization_Linear_Density_Phase}
\end{equation}
where we defined,
\begin{equation}
\begin{aligned}c & =\sqrt{\left(v_{\mathrm{F}}+\frac{g_{4}}{2\pi}\right)^{2}-\left(\frac{g_{2}}{2\pi}\right)^{2}}\,,\\
K & =\sqrt{\frac{v_{\mathrm{F}}+\frac{g_{4}}{2\pi}-\frac{g_{2}}{2\pi}}{v_{\mathrm{F}}+\frac{g_{4}}{2\pi}+\frac{g_{2}}{2\pi}}}\,,
\end{aligned}
\label{eq:Luttinger_parameters_Fermions}
\end{equation}
Note that $c\approx v_{\mathrm{F}}$ and $K\approx1$ because the
initial fermions are weakly interacting. We also defined the new source
fields coupled to $\theta$ and $\phi$ in terms of the ones coupled
to $\chi_{\pm}$ as,
\[
u_{\theta,\phi}=\frac{u_{+}\pm u_{-}}{2}
\]
Relation (\ref{eq:Movers_Chiral_relation}) is now updated to,$^{\textrm{\ref{fn:KeldyshFactor2}}}$
\begin{equation}
\begin{aligned} & \bar{\psi}_{+}\psi_{+}+\bar{\psi}_{-}\psi_{-}\longleftrightarrow\frac{1}{\pi}\partial_{x}\theta\,.\\
 & \bar{\psi}_{+}\psi_{+}-\bar{\psi}_{-}\psi_{-}\longleftrightarrow\frac{1}{\pi}\partial_{x}\phi\,.
\end{aligned}
\label{eq:Movers_Density-Phase_relations}
\end{equation}
The first line is the sum of the densities of right and left movers,
which defines the non-homogeneous part of the density. Remembering
that we are measuring the non-homogeneous part of the density from
its homogeneous part, we find that $\frac{1}{\pi}\partial_{x}\theta$
measures the density fluctuations (in the low-energy limit). The term
$\phi^{\alpha}\partial_{t}\left(\frac{1}{\pi}\partial_{x}\theta_{\alpha}\right)$
in partition function (\ref{eq:Fermion_Bosonization_Linear_Density_Phase})
is the term of the Legendre transform that relates the Lagrangian
and the Hamiltonian, such as $p\dot{x}$ in $L=p\dot{x}-H$ in quantum
mechanics. Then, in the same way $p$ is conjugate to $x$, we deduce
that $\phi$ is the phase conjugate to the density $\frac{1}{\pi}\partial_{x}\theta$.
We conclude that we have obtained the hydrodynamic formalism given
by the Luttinger liquid Hamiltonian, Eq. (\ref{eq:Luttinger-liquid_Hamiltonian_plus_Corrections}),
without the non-linear corrections and the constant energy term.\footnote{For the comparison, note that there is an additional factor 2 multiplying
the Lagrangian in the Keldysh formalism.} Therefore, the the parameter $c$ and $K$ in Eq. (\ref{eq:Luttinger_parameters_Fermions})
are the speed of sound and the Luttinger parameter. From the second
line of Eq. (\ref{eq:Movers_Density-Phase_relations}) and Eq. (\ref{eq:Density-Velocity_phases})
we also find that the low-energy velocity field in terms of right
and left movers is $v=\frac{\pi}{m}(\bar{\psi}_{+}\psi_{+}-\bar{\psi}_{-}\psi_{-})$.

\subsection{Non-linear corrections}

In the last subsection we derived the linear bosonization, corresponding
to the first two diagrams of Fig. \ref{fig:Vertex_Corrections}. The
first diagram was omitted by measuring the density from its homogeneous
value and the second diagram led to a quadratic low-energy theory
of hydrodynamic excitations. As part of the original contributions
of this work, in this sub-section we derive the first non-linear correction
to linear bosonization, the three-leg diagram in Fig. \ref{fig:Vertex_Corrections},
within the Keldysh functional integral. We show that perturbation
theory in the spectrum curvature works well to derive the three-leg
correction. We check this result using a more rigorous approach based
on the Matsubara formalism, derived in Appendix \ref{sec:N-loop-reduction-formula},
where we also derive the four-leg correction and formulate a conjecture
for all other terms.

The non-linear part of series (\ref{eq:Bosonization_Series}), given
by the sum with $n\geq3$, that we denote as $M_{\eta}[\varrho_{\eta}^{\alpha}]$,
can be reformulated using the knowledge of the polarisation diagram.
We split linear and non-linear parts of $\mathrm{Tr}\log(1-G_{\eta}\circ\varrho_{\eta})$
in (\ref{eq:Partition_Function_chi_Fourier_Transform}),
\[
\tilde{Z}_{\eta}[\chi_{\eta}]=\int\mathcal{D}\varrho_{\eta}e^{\mathrm{i}\int\mathrm{d}r\,\frac{1}{\pi}\chi_{\eta}^{\prime\alpha}\varrho_{\eta\alpha}-\frac{\mathrm{i}}{2}\int\mathrm{d}r\mathrm{d}r'\,\varrho_{\eta}^{\alpha}(r)\Pi_{\eta}^{\alpha\beta}(r-r')\varrho_{\eta}^{\beta}(r')-M_{\eta}[\varrho_{\eta}^{\alpha}]}\,,
\]
and express the non-linear term, $M_{\eta}[\varrho_{\eta}^{\alpha}]$,
in terms of the functional Fourier transform field, $\chi_{\eta}^{\prime\alpha}$,
by substituting $\varrho_{\eta}^{\alpha}$ with $-\mathrm{i}\pi\frac{\delta}{\delta\chi'_{\eta\alpha}}$
and moving $M_{\eta}[-\mathrm{i}\pi\frac{\delta}{\delta\chi'_{\eta\alpha}}]$
in front of the integral,
\[
\begin{aligned}\tilde{Z}_{\eta}[\chi_{\eta}] & =e^{-M_{\eta}[-\mathrm{i}\pi\frac{\delta}{\delta\chi'_{\alpha}}]}\int\mathcal{D}\varrho_{\eta}e^{\mathrm{i}\int\mathrm{d}r\,\frac{1}{\pi}\chi_{\eta\alpha}^{\prime}\varrho_{\eta}^{\alpha}-\frac{\mathrm{i}}{2}\int\mathrm{d}r\mathrm{d}r'\,\varrho_{\eta}^{\alpha}(r)\Pi_{\eta}^{\alpha\beta}(r-r')\varrho_{\eta}^{\beta}(r')}\\
 & =e^{-M_{\eta}[-\mathrm{i}\pi\frac{\delta}{\delta\chi'_{\eta\alpha}}]}e^{\mathrm{i}\int\mathrm{d}r\,\frac{1}{2}\chi_{\eta}^{\alpha}D_{0,\eta}^{-1}\chi_{\eta\alpha}}\\
 & =e^{\mathrm{i}\int\mathrm{d}r\,\frac{1}{2}\chi_{\eta}^{\alpha}D_{0,\eta}^{-1}\chi_{\eta\alpha}}\left[e^{-\mathrm{i}\int\mathrm{d}r\,\frac{1}{2}\chi_{\eta}^{\alpha}D_{0,\eta}^{-1}\chi_{\eta\alpha}}e^{-M_{\eta}[-\mathrm{i}\pi\frac{\delta}{\delta\chi'_{\eta\alpha}}]}e^{\mathrm{i}\int\mathrm{d}r\,\frac{1}{2}\chi_{\eta}^{\alpha}D_{0,\eta}^{-1}\chi_{\eta\alpha}}\right]\,,
\end{aligned}
\]
where in the second line we integrated over $\varrho_{\eta}$ and
used the linear bosonization results (\ref{eq:Partition_Function_Chi_2})
and (\ref{eq:Chi_Propagator_no_g}) and in the third line we multiplied
and divided by $e^{\mathrm{i}\int\mathrm{d}r\,\frac{1}{2}\chi_{\eta}^{\alpha}D_{0,\eta}^{-1}\chi_{\eta\alpha}}$.
Expanding the first term in a Taylor series in $-\mathrm{i}\pi\frac{\delta}{\delta\chi'_{\alpha}}$
and inserting $1=e^{\mathrm{i}\int\mathrm{d}r\,\frac{1}{2}\chi_{\eta}^{\alpha}D_{0,\eta}^{-1}\chi_{\eta\alpha}}e^{-\mathrm{i}\int\mathrm{d}r\,\frac{1}{2}\chi_{\eta}^{\alpha}D_{0,\eta}^{-1}\chi_{\eta\alpha}}$
between each derivative, we obtain the shift $-\mathrm{i}\pi\frac{\delta}{\delta\chi'_{\alpha}}\rightarrow-\eta\partial_{\eta}\chi_{\eta}^{\alpha}-\mathrm{i}\pi\frac{\delta}{\delta\chi'_{\alpha}}$,
where we defined,
\[
\partial_{\eta}=\partial_{t}+\eta v_{\mathrm{F}}\partial_{x}\,,
\]
that leads to,
\begin{figure}
\centering
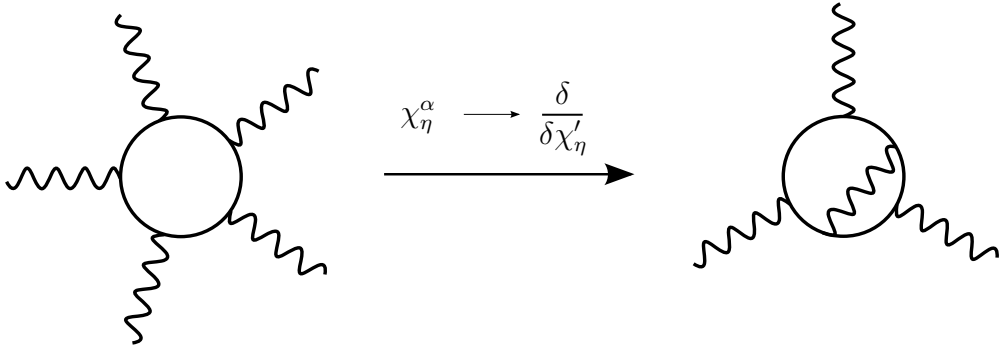

\caption[Example of contraction of two legs of the five-loop.]{Example of contraction of two legs of the five-loop caused by the
replacement of $\sim\chi_{\eta}^{\alpha}$ by $\sim\frac{\delta}{\delta\chi'_{\eta}}$
inside $M_{\eta}$ in partition function (\ref{eq:Partition_Function_Non-linear_Bosonization}).
The contraction gives a correction to the semiclassical three-loop
shown in Fig. \ref{fig:Vertex_Corrections}.\label{fig:3loop_Radiative_Correction}}
\end{figure}
 
\[
\tilde{Z}_{\eta}[\chi_{\eta}]=e^{\mathrm{i}\int\mathrm{d}r\,\frac{1}{2}\chi_{\eta}^{\alpha}D_{0,\eta}^{-1}\chi_{\eta\alpha}-M_{\eta}[-\eta\partial_{\eta}\chi_{\eta}^{\alpha}-\mathrm{i}\pi\frac{\delta}{\delta\chi'_{\eta\alpha}}]}\,.
\]
Inserting this result in the first line of Eq. (\ref{eq:Partition_Function_Keldysh_Fermions_Full})
and using Eq. (\ref{eq:Chi_Propagator}) we find, 
\begin{equation}
Z[u]=\int\mathcal{D}\chi_{\pm}e^{\mathrm{i}\int\mathrm{d}r\,\left[\frac{1}{2}\chi_{\eta}^{\alpha}D_{\eta\eta'}^{-1}\chi_{\eta'\alpha}-\frac{1}{\pi}\partial_{x}\chi_{\eta}^{\alpha}u_{\eta\alpha}\right]-\sum_{\eta}M_{\eta}[-\eta\partial_{\eta}\chi_{\eta}^{\alpha}-\mathrm{i}\pi\frac{\delta}{\delta\chi'_{\eta\alpha}}]}\,.\label{eq:Partition_Function_Non-linear_Bosonization}
\end{equation}
This partition function contains two contributions in the exponent:
the first term is the linear contribution found previously and the
second term is the non-linear contribution. The argument of $M_{\eta}$
is the sum of two terms, $\sim\chi_{\eta}^{\alpha}$ and $\sim\frac{\delta}{\delta\chi'_{\eta\alpha}}$.
Let us explore what this means by considering the fifth power in the
expansion of $M_{\eta}$ from Eq. (\ref{eq:Bosonization_Series}).
If the derivative term, $\sim\frac{\delta}{\delta\chi'_{\eta\alpha}}$,
were not present, we would have a term proportional to the product
of five fields $\chi_{\eta}\chi_{\eta}\chi_{\eta}\chi_{\eta}\chi_{\eta}$,
where we omitted variables and indices for brevity. The presence of
$\sim\frac{\delta}{\delta\chi'_{\eta\alpha}}$ summed to $\sim\chi_{\eta}^{\alpha}$
in the power series, means that we consider terms where $\sim\chi_{\eta}^{\alpha}$
is replaced by $\sim\frac{\delta}{\delta\chi'_{\eta\alpha}}$, such
as $\sim\chi_{\eta}\chi_{\eta}\chi_{\eta}\frac{\delta}{\delta\chi'_{\eta\alpha}}\chi_{\eta}\sim\chi_{\eta}\chi_{\eta}\chi_{\eta}$,
where in the second step we acted with the derivative. Following this
example, we see that replacing $\sim\chi_{\eta}^{\alpha}$ with $\sim\frac{\delta}{\delta\chi'_{\eta\alpha}}$
in the power series expansion of $M_{\eta}$ amounts to reducing by
two the number of fields in each term. By doing the explicit calculation
it is easy to show that this generates contractions between pairs
of fields $\chi_{\eta}^{\alpha}$, as shown in Fig. \ref{fig:3loop_Radiative_Correction}
in the case of five fields. Because contractions between pairs of
fields generate quantum corrections \cite{ZinnJustin}, neglecting
the derivative in the argument of $M_{\eta}$ amounts to discarding
quantum corrections, leading to a semiclassical approximation. Moreover,
as we show in Appendix \ref{sec:N-loop-reduction-formula}, the quantum
corrections are small for small momenta and, because we are considering
the case $q\ll mc$, we can neglect them by dropping the term $\sim\frac{\delta}{\delta\chi'_{\eta}}$,
\[
M_{\eta}[-\eta\partial_{\eta}\chi_{\eta}^{\alpha}-\mathrm{i}\pi\frac{\delta}{\delta\chi'_{\eta\alpha}}]\;\overset{\textrm{semiclassical}}{\approx}\;M_{\eta}[-\eta\partial_{\eta}\chi_{\eta}^{\alpha}]\,.
\]

In the semiclassical approximation, we start exploring the first non-linear
correction in $M_{\eta}$, that is, the three-leg vertex, whose additional
term to the action is,
\begin{equation}
S_{\mathrm{nl}}[\chi_{\eta}^{\alpha}]=\frac{\eta}{3}\int\mathrm{d}r_{1}\mathrm{d}r_{2}\mathrm{d}r_{3}\Gamma_{3,\eta}^{\alpha_{1}\alpha_{2}\alpha_{3}}(r_{1},r_{2},r_{3})\partial_{\eta}\chi_{\eta}^{\alpha_{1}}(r_{1})\partial_{\eta}\chi_{\eta}^{\alpha_{2}}(r_{2})\partial_{\eta}\chi_{\eta}^{\alpha_{3}}(r_{3})\,.\label{eq:Action_three-leg}
\end{equation}
More specifically, since the main contribution in the semiclassical
approximation is given by the equation of $\chi_{\eta}^{\mathrm{cl}}$
\cite{Kamenev2011}, we need the product of two classical fields and
one quantum field,
\begin{equation}
S_{\mathrm{nl}}[\chi_{\eta}^{\alpha}]=\frac{\eta}{3}\int\mathrm{d}r_{1}\mathrm{d}r_{2}\mathrm{d}r_{3}\Gamma_{3,\eta}^{\mathrm{cl}\mathrm{cl}\mathrm{q}}(r_{1},r_{2},r_{3})\partial_{\eta}\chi_{\eta}^{\mathrm{cl}}(r_{1})\partial_{\eta}\chi_{\eta}^{\mathrm{cl}}(r_{2})\partial_{\eta}\chi_{\eta}^{\mathrm{q}}(r_{3})\,,\label{eq:Action_three-leg_clclq}
\end{equation}
where the vertex is,
\[
\begin{aligned}\Gamma_{3,\eta}^{\mathrm{cl}\mathrm{cl}\mathrm{q}}(r_{1},r_{2},r_{3}) & =G_{\eta}^{\mathrm{R}}(r_{1}-r_{2})G_{\eta}^{\mathrm{K}}(r_{2}-r_{3})G_{\eta}^{\mathrm{R}}(r_{3}-r_{1})\\
 & +G_{\eta}^{\mathrm{K}}(r_{1}-r_{2})G_{\eta}^{\mathrm{A}}(r_{2}-r_{3})G_{\eta}^{\mathrm{R}}(r_{3}-r_{1})\\
 & +G_{\eta}^{\mathrm{A}}(r_{1}-r_{2})G_{\eta}^{\mathrm{A}}(r_{2}-r_{3})G_{\eta}^{\mathrm{K}}(r_{3}-r_{1})\,.
\end{aligned}
\]
To calculate the vertex we Fourier transform,
\[
\begin{aligned}-\frac{\mathrm{i}\eta}{3}\int\mathrm{d}Q_{1}\mathrm{d}Q_{2}\,\Gamma_{3,\eta}^{\mathrm{cl}\mathrm{cl}\mathrm{q}}(Q_{1},Q_{2}) & (\omega_{1}-\eta cq_{1})\chi_{\eta}^{\mathrm{cl}}(Q_{1})(\omega_{2}-\eta cq_{2})\chi_{\eta}^{\mathrm{cl}}(Q_{2})\\
 & \times(\omega_{1}+\omega_{2}-\eta c(q_{1}+q_{2}))\chi_{\eta}^{\mathrm{q}}(-Q_{1}-Q_{2})\,,
\end{aligned}
\]
where we use the shorthand notation $Q_{i}=(q_{i},\omega_{i})$ and,
\[
\begin{aligned}\Gamma_{3,\eta}^{\mathrm{cl}\mathrm{cl}\mathrm{q}}(Q_{1},Q_{2})=\int\frac{\mathrm{d}k}{2\pi}\frac{\mathrm{d}\varepsilon}{2\pi} & \left[G_{\eta}^{\mathrm{R}}(K)G_{\eta}^{\mathrm{K}}(K+Q_{1})G_{\eta}^{\mathrm{R}}(K+Q_{1}+Q_{2})\right.\\
 & +G_{\eta}^{\mathrm{K}}(K)G_{\eta}^{\mathrm{A}}(K+Q_{1})G_{\eta}^{\mathrm{R}}(K+Q_{1}+Q_{2})\\
 & \left.+G_{\eta}^{\mathrm{A}}(K)G_{\eta}^{\mathrm{A}}(K+Q_{1})G_{\eta}^{\mathrm{K}}(K+Q_{1}+Q_{2})\right]\,.
\end{aligned}
\]
Integrating over $\varepsilon$ we have,
\[
\begin{aligned}\Gamma_{3,\eta}^{\mathrm{cl}\mathrm{cl}\mathrm{q}}(Q_{1},Q_{2}) & =\\
=-\mathrm{i}\int\frac{\mathrm{d}k}{2\pi} & \left[-\frac{\tanh(\varepsilon_{\eta}(k+q_{1})/2T)}{[\omega_{1}-\varepsilon_{\eta}(k+q_{1})+\varepsilon_{\eta}(k)][\omega_{2}-\varepsilon_{\eta}(k+q_{1}+q_{2})+\varepsilon_{\eta}(k+q_{1})]}\right.\\
 & +\frac{\tanh(\varepsilon_{\eta}(k)/2T)}{[\omega_{1}-\varepsilon_{\eta}(k+q_{1})+\varepsilon_{\eta}(k)][\omega_{1}+\omega_{2}-\varepsilon_{\eta}(k+q_{1}+q_{2})+\varepsilon_{\eta}(k)]}\\
 & \left.+\frac{\tanh(\varepsilon_{\eta}(k+q_{1}+q_{2})/2T)}{[\omega_{2}-\varepsilon_{\eta}(k+q_{1}+q_{2})+\varepsilon_{\eta}(k+q_{1})][\omega_{1}+\omega_{2}-\varepsilon_{\eta}(k+q_{1}+q_{2})+\varepsilon_{\eta}(k)]}\right]\,,
\end{aligned}
\]
where $\varepsilon_{\eta}(k)=\eta ck+k^{2}/2m$. We split the first
term using the identity $1/ab=(1/a+1/b)/(a+b)$ and get
\[
\begin{aligned}\Gamma_{3,\eta}^{\mathrm{cl}\mathrm{cl}\mathrm{q}}(Q_{1},Q_{2}) & =\\
=\mathrm{i}\int\frac{\mathrm{d}k}{2\pi} & \left[\frac{\tanh(\varepsilon_{\eta}(k+q_{1})/2T)-\tanh(\varepsilon_{\eta}(k)/2T)}{[\omega_{1}-\varepsilon_{\eta}(k+q_{1})+\varepsilon_{\eta}(k)][\omega_{1}+\omega_{2}-\varepsilon_{\eta}(k+q_{1}+q_{2})+\varepsilon_{\eta}(k)]}\right.\\
 & \left.-\frac{\tanh(\varepsilon_{\eta}(k+q_{1}+q_{2})/2T)-\tanh(\varepsilon_{\eta}(k+q_{1})/2T)}{[\omega_{2}-\varepsilon_{\eta}(k+q_{1}+q_{2})+\varepsilon_{\eta}(k+q_{1})][\omega_{1}+\omega_{2}-\varepsilon_{\eta}(k+q_{1}+q_{2})+\varepsilon_{\eta}(k)]}\right]\,.
\end{aligned}
\]
We expand the integral up to the second power in the small spectrum
curvature, $m^{-1}$, which, as we saw previously, is equivalent to
a small momentum expansion, $q\ll k_{\mathrm{F}}$. This expansion
is convergent term by term and the first two are calculated more rigorously
in Appendix \ref{sec:N-loop-reduction-formula} using the Matsubara
formalism. As the vertices are symmetric for the exchange of external
legs, we consider only their symmetrised part. The zero order symmetrised
term in $1/m$ is zero, consistent with the Dzyaloshinski-Larkin theorem.
Up to the third order we have,
\[
\Gamma_{3,\eta}^{\mathrm{cl}\mathrm{cl}\mathrm{q}}(Q_{1},Q_{2})=-\frac{\mathrm{i}\eta}{2\pi m}\frac{q_{1}q_{2}(q_{1}+q_{2})}{(\omega_{1}-\eta cq_{1})(\omega_{2}-\eta cq_{2})(\omega_{1}+\omega_{2}-\eta c(q_{1}+q_{2}))}+\mathcal{O}\left(m^{-3}\right)\,.
\]
This result is already symmetric and does not need symmetrisation.
We also multiplied by a factor 3, as there are three ways of choosing
the quantum field in Eq. (\ref{eq:Action_three-leg}). We note the
absence of the second order term in $m^{-1}$. Then, Eq. (\ref{eq:Action_three-leg_clclq})
becomes,
\[
\mathrm{i}S_{\mathrm{nl}}[\chi_{\eta}^{\alpha}]\approx\frac{1}{2\pi m}\int\mathrm{d}Q_{1}\mathrm{d}Q_{2}\,q_{1}\chi_{\eta}^{\mathrm{cl}}(Q_{1})q_{2}\chi_{\eta}^{\mathrm{cl}}(Q_{2})(q_{1}+q_{2})\chi_{\eta}^{\mathrm{q}}(-Q_{1}-Q_{2})\,.
\]
Taking the Fourier transform we have,
\begin{equation}
S_{\mathrm{nl}}[\chi_{\eta}^{\alpha}]=-\frac{1}{2\pi m}\int\mathrm{d}r\,\left(\chi_{\eta}^{\prime\mathrm{cl}}(r)\right)^{2}\chi_{\eta}^{\prime\mathrm{q}}(r)\,,\label{eq:Non-Linear_Chiral_Correction}
\end{equation}
and adding this term to the quadratic part of the action in partition
function (\ref{eq:Partition_Function_chi}), we have,
\[
Z[u]=\int\mathcal{D}\chi_{\pm}\exp\mathrm{i}\int\mathrm{d}r\,\left[\frac{1}{2}\chi_{\eta}^{\alpha}D_{\eta\eta'}^{-1}\chi_{\eta'\alpha}-\frac{1}{2\pi m}\left(\chi_{\eta}^{\prime\mathrm{cl}}\right)^{2}\chi_{\eta}^{\prime\mathrm{q}}-\frac{1}{\pi}\chi_{\eta}^{\prime\alpha}u_{\eta\alpha}\right]\,.
\]
Moreover, in the $\theta-\phi$ representation the non-linear term
(\ref{eq:Non-Linear_Chiral_Correction}) becomes,
\begin{equation}
S_{\mathrm{nl}}[\theta^{\alpha},\phi^{\alpha}]=-\frac{1}{\pi m}\int\mathrm{d}r\,\left\{ \left[\left(\theta^{\prime\mathrm{cl}}(r)\right)^{2}+\left(\phi^{\prime\mathrm{cl}}(r)\right)^{2}\right]\theta^{\prime\mathrm{q}}(r)+2\theta^{\prime\mathrm{cl}}(r)\phi^{\prime\mathrm{cl}}(r)\phi^{\prime\mathrm{q}}(r)\right\} \,.\label{eq:Fermion_Bosonization_Non-Linear_Density_Phase}
\end{equation}
For the chiral fields, $\chi_{\pm}$, and the phase fields, $\theta$
and $\phi$, the mass curvature, $m^{-1}$, becomes the parameter
that control the interaction.

The three-leg correction, being proportional to the mass curvature,
is smaller and smaller the closer we are to the Fermi points or the
smaller the cut-off, $q_{0}$, is, as can be understood by Fig. \ref{fig:Linearization_Spectrum}.
This fact, however, does not exclude the presence of higher non-linear
terms, such as the four-leg diagram in Fig. \ref{fig:Vertex_Corrections}.
To exclude them, we have to show that higher non-linear terms are
smaller than the three-leg term in the mass curvature. In support
of this fact, at zero temperature, we have the Dzyaloshinski-Larking
theorem discussed previously. But, we can have a better feel by using
dimensional analysis. By rescaling the $n$-leg contribution, Eq.
(\ref{eq:n-vertex}), by the characteristic energy and momentum of
the system, $mv_{\mathrm{F}}^{2}$ and $mv_{\mathrm{F}}$, we find
that the n-loop behaves as $\widetilde{\Gamma}_{n}\sim(mv_{\mathrm{F}}^{2})^{-(n-2)}$,
where we defined $\widetilde{\Gamma}_{n,\eta}$ as the symmetrised
$n$-leg vertex for a correct treatment of n-loops, as seen in Appendix
\ref{sec:N-loop-reduction-formula}.\footnote{Note that the characteristic energy and momentum coming from the fermionic
interaction terms, $\sim g_{2}$ and $\sim g_{4}$, do not participate
in the evaluation of the n-loop corrections.} In turn, we have $\widetilde{\Gamma}_{n+1}/\widetilde{\Gamma}_{n}\sim m^{-1}$,
that means that for a smaller spectrum curvature, $m^{-1}$, higher
n-loop corrections become increasingly weaker. At finite but small
temperatures, $T\ll mv_{\mathrm{F}}^{2}$, the result must be consistent
with the one at zero temperature in the limit of $T\rightarrow0$,
which implies that the thermal corrections to the n-loop can only
have positive powers of the temperature, $\widetilde{\Gamma}_{n}\sim T^{\ell},\;\ell\geq1$.
Putting the zero and finite temperature dimensional analysis together,
we conclude that the n-loop can only contain terms of the type,
\[
\widetilde{\Gamma}_{n}\sim(mv_{\mathrm{F}}^{2})^{-(n-2)}\left(\frac{T}{mv_{\mathrm{F}}^{2}}\right)^{\ell},\;\ell\geq0
\]
and the term with $\ell=0$ corresponds to the zero temperature contribution.
So, in first approximation in $T/mv_{\mathrm{F}}^{2}\ll1$, we have
a small temperature extension of the Dzyaloshinski-Larking theorem,
\begin{enumerate}
\item $\widetilde{\Gamma}_{n}=\mathcal{O}(m^{-(n-2)})$ for $m^{-1}\rightarrow0$;
\item $\widetilde{\Gamma}_{n}$ does not depend on temperature at the lowest
order in $m^{-1}$, that is, $m^{-(n-2)}$.
\end{enumerate}
We verify this result in Appendix \ref{sec:N-loop-reduction-formula}
for the three-, four- and five-loop and find $\widetilde{\Gamma}_{3}=\mathcal{O}(m^{-1})$,
$\widetilde{\Gamma}_{4}=\mathcal{O}(m^{-2})$ and $\widetilde{\Gamma}_{5}=\mathcal{O}(m^{-3})$.
Moreover, in the expansion in powers of $m^{-1}$, the first non-zero
terms, respectively proportional to $\widetilde{\Gamma}_{3}\sim m^{-1}$,
$\widetilde{\Gamma}_{4}\sim m^{-2}$ and $\widetilde{\Gamma}_{5}\sim m^{-3}$,
are temperature independent.

Within this argument, successive terms of series (\ref{eq:Bosonization_Series}),
given by fermionic loops with an increasing number of phononic external
legs as in Fig. \ref{fig:Vertex_Corrections}, are smaller and smaller
in the spectrum curvature or closer to the Fermi points, justifying
the dropping of loops with more than three legs for small momenta.

\section{Boson hydrodynamics}

In the previous section we derived a low-energy hydrodynamic description
of weakly interacting fermionic particles in one dimension using the
technique of functional bosonization. In this section, we derive the
low-energy hydrodynamic description of bosonic particles with repulsive
contact interaction in the limits of weak and strong interaction.

The Lagrangian of the system is,
\begin{equation}
L\left[\bar{\varphi},\varphi\right]=\int\mathrm{d}x\,\left[\bar{\varphi}\left(\mathrm{i}\partial_{t}+\frac{1}{2m}\partial_{x}^{2}+\mu\right)\varphi-\frac{g}{2}(\bar{\varphi}\varphi)^{2}\right]\,,\label{eq:Lagrangian_Interacting_Bosons}
\end{equation}
where $\varphi(x,t)$ is the bosonic particle field, $m$ is the mass
of the particle, $\mu$ the chemical potential and $g$ the interaction
strength of the density-density interaction. The system described
by Lagrangian (\ref{eq:Lagrangian_Interacting_Bosons}) is homogeneous
as it is not subject to any external potential. Then, we consider
the mean homogeneous density of the system, $n$, and distinguish
two cases \cite{Dalfovo1999,PitaevskiiStringari}:
\begin{enumerate}
\item Weak interaction or high density, $mg\ll n$;
\item Strong interaction or low density, $mg\gg n$ (Tonks-Girardeau gas);\footnote{Reintroducing $\hbar$, the left and right terms in the inequalities
read $\hbar n$ and $mg/\hbar$. Setting $\hbar=1$, $n$ and $mg$
have the dimensions of a momentum.}
\end{enumerate}
We start from the simpler case of weak interaction.

\subsection{Weak interactions\label{subsec:Weak-interactions}}

For weak interactions or high density, it is useful to introduce the
healing length, $\xi=\frac{1}{2\sqrt{mgn}}$, the characteristic length
of the interacting system. The number of particles within a healing
length is $n\xi=\frac{1}{2}\sqrt{\frac{n}{mg}}\propto\sqrt{n}$ and
increases with density. At high densities, there are many particles
in a healing length, satisfying the criterion for the applicability
of the mean field theory \cite{PitaevskiiStringari}. Then, the dominant
contribution to the functional integral is the mean field solution
$\varphi=\bar{\varphi}=\sqrt{n}$ and minimisation of the energy $H\left[\bar{\varphi},\varphi\right]=\int\mathrm{d}x\,\bar{\varphi}\mathrm{i}\partial_{t}\varphi-L\left[\bar{\varphi},\varphi\right]$
gives the condition,
\begin{equation}
\mu=gn\,.\label{eq:Chemical_Potential_Weakly_Interacting}
\end{equation}
Now, we consider small fluctuations of the bosonic field, $\varphi$,
around its mean value, $\sqrt{n}$. To do this, we write the bosonic
field using the phase-density representation,

\begin{equation}
\varphi(x,t)=\sqrt{n+\rho(x,t)}e^{i\phi(x,t)}\,,\label{eq:Phase-Density_representatioin}
\end{equation}
where $\rho$ and $\phi$ are small density and phase fluctuations.
Substituting phase-density representation (\ref{eq:Phase-Density_representatioin})
in Lagrangian (\ref{eq:Lagrangian_Interacting_Bosons}) we find,
\begin{equation}
L\left[\rho,\phi\right]=\int\mathrm{d}x\,\left[-\rho\partial_{t}\phi-\frac{n}{2m}\left(\partial_{x}\phi\right)^{2}-\frac{g}{2}\rho^{2}-\frac{1}{2m}\rho\left(\partial_{x}\phi\right)^{2}-\frac{1}{8m}\frac{\left(\partial_{x}\rho\right)^{2}}{n+\rho}\right]\,,\label{eq:Phase-Density_representation_weakly_interacting_bosons}
\end{equation}
where we omitted constant and boundary terms. The last term is called
quantum pressure and accounts for quantum effects. As we consider
small density fluctuations, $\rho$, that is, fluctuations with size
greater than the healing length, the quantum pressure can be neglected
\cite{PitaevskiiStringari}. Comparing this hydrodynamic Lagrangian
with Hamiltonian (\ref{eq:Hydrodynamic_Hamiltonian}), the compressibility
of the system is $\kappa=1/gn^{2}$ and, as one would expect, it is
harder to compress the system when the repulsion is stronger or the
density is higher. Expressing the density in term of the phase $\theta$,
as $\rho=\frac{1}{\pi}\partial_{x}\theta$, the Lagrangian becomes,
\[
L\left[\theta,\phi\right]=\int\mathrm{d}x\,\left[-\frac{1}{\pi}\partial_{x}\theta\partial_{t}\phi-\frac{n}{2m}\left(\partial_{x}\phi\right)^{2}-\frac{g}{2\pi^{2}}(\partial_{x}\theta)^{2}-\frac{1}{2m\pi}\partial_{x}\theta\left(\partial_{x}\phi\right)^{2}\right]\,,
\]
where the first three terms are the linear part and the last term
the non-linear part. Integrating by parts we have,
\begin{equation}
L\left[\theta,\phi\right]=\int\mathrm{d}x\,\frac{1}{2}\left(\begin{array}{c}
\theta\\
\phi
\end{array}\right)^{\mathrm{T}}\left(\begin{array}{cc}
\frac{c}{\pi K}\partial_{x}^{2} & \frac{1}{\pi}\partial_{t}\partial_{x}\\
\frac{1}{\pi}\partial_{t}\partial_{x} & \frac{cK}{\pi}\partial_{x}^{2}
\end{array}\right)\left(\begin{array}{c}
\theta\\
\phi
\end{array}\right)-\int\mathrm{d}x\,\left[\frac{1}{2m\pi}\partial_{x}\theta\left(\partial_{x}\phi\right)^{2}\right]\,,\label{eq:Bosonized_boson_Action}
\end{equation}
where,
\begin{equation}
\begin{aligned}c= & \sqrt{\frac{ng}{m}}\,,\\
\frac{K}{\pi}= & \sqrt{\frac{n}{mg}}\,,
\end{aligned}
\label{eq:Luttinger_parameters_Weakly_interacting_bosons}
\end{equation}
are the speed of sound and the Luttinger parameter. Finally, we consider
the action $S\left[\theta,\phi\right]=\int_{\rightleftarrows}\mathrm{d}tL\left[\theta,\phi\right]$
and perform a Keldysh rotation to find the hydrodynamic action: the
linear part is the same as the fermionic one, Eq. (\ref{eq:Fermion_Bosonization_Linear_Density_Phase});
however, the non-linear part does not contain the term cubic in $\partial_{x}\theta$,
present in the fermionic case, Eq. (\ref{eq:Fermion_Bosonization_Non-Linear_Density_Phase}).

\subsection{Strong interactions}

In the case of strong interaction or low density, the system becomes
a Tonks-Girardeau gas of hard-core bosons \cite{Girardeau60,Lieb_Liniger_1963}.
The partition function can be written as
\[
\begin{aligned}Z & =\int\mathcal{D}[\bar{\varphi},\varphi]e^{\mathrm{i}\int\mathrm{d}t\mathrm{d}x\,\left[\bar{\varphi}\left(\mathrm{i}\partial_{t}+\frac{1}{2m}\partial_{x}^{2}+\mu\right)\varphi-\frac{g}{2}\bar{\varphi}^{2}\varphi^{2}\right]}\\
 & =\int\mathcal{D}[\bar{\varphi},\varphi]e^{\mathrm{i}\int\mathrm{d}t\mathrm{d}x\,\left[\bar{\varphi}\left(\mathrm{i}\partial_{t}+\frac{1}{2m}\partial_{x}^{2}+\mu\right)\varphi\right]}\int\mathcal{D}[\bar{\xi},\xi]e^{\mathrm{i}\int\mathrm{d}t\mathrm{d}x\,\left[\frac{2}{g}\bar{\xi}\xi+\xi\bar{\varphi}^{2}+\bar{\xi}\varphi^{2}\right]}\,,
\end{aligned}
\]
where in the second line we introduced the field $\xi$ through a
Hubbard-Stratonovich transformation. The last term in the limit $g\gg n/m$
becomes,
\[
\begin{aligned}\int\mathcal{D}[\bar{\xi},\xi]e^{\mathrm{i}\int\mathrm{d}t\mathrm{d}x\,\left[\frac{2}{g}\bar{\xi}\xi+\xi\bar{\varphi}\bar{\varphi}+\bar{\xi}\varphi\varphi\right]} & \approx\int\mathcal{D}[\bar{\xi},\xi]e^{\mathrm{i}\int\mathrm{d}t\mathrm{d}x\,\left[\xi\bar{\varphi}\bar{\varphi}+\bar{\xi}\varphi\varphi\right]}=\\
 & =\delta(\mathrm{Re}(\varphi^{2}))\delta(\mathrm{Im}(\varphi^{2}))\\
 & =\delta(\bar{\varphi}^{2})\delta(\varphi^{2})\,,
\end{aligned}
\]
which imposes the constraint $\varphi(x)\varphi(x)=\bar{\varphi}(x)\bar{\varphi}(x)=0$,
that is, two bosons cannot be at the same position.\footnote{Note that, alternatively, the interaction term can be split as $\int\mathcal{D}\varrho e^{\mathrm{i}\int\mathrm{d}t\mathrm{d}x\,\left[\frac{1}{2g}\varrho^{2}+\varrho\bar{\varphi}\varphi\right]}$,
which, in the limit $mg\gg n$, would lead to the constraint $\bar{\varphi}(x)\varphi(x)=0$.
However, this imposes that the density is zero or, in other words,
that there are no particles in the system. Having no particles in
the system implies that no two bosons are at the same position, but
the constraint that two bosons cannot be at the same position does
not imply that there are no particles in the system. Therefore, for
a more general constraint we need the weaker condition, that is, two
bosons cannot be at the same position.} The partition function simplifies to
\begin{equation}
Z=\int\mathcal{D}[\bar{\varphi},\varphi]\delta(\bar{\varphi}^{2})\delta(\varphi^{2})e^{\mathrm{i}\int\mathrm{d}t\mathrm{d}x\,\left[\bar{\varphi}\left(\mathrm{i}\partial_{t}+\frac{1}{2m}\partial_{x}^{2}+\mu\right)\varphi\right]}\,.\label{eq:Partition_Function_Hard-Core_Bosons}
\end{equation}
The constraint means that the boson fields anti-commute at the same
position, $[\varphi(x),\varphi(x)]_{-}=2\varphi(x)\varphi(x)=0$,
and commute at different positions, $[\varphi(x),\varphi(x')]_{+}=0,\;x\neq x'$.
The anti-commutation property suggests to change the bosonic complex
fields $\varphi(x)$ and $\bar{\varphi}(x)$ to the fermionic Grassmann
fields $\psi(x)$ and $\bar{\psi}(x)$. Through the anti-commutativity
of the fermions, the constraint $\varphi(x)\varphi(x)=\psi(x)\psi(x)=0$
is automatically satisfied. However, the change of variables must
respect the commutativity $[\varphi(x),\varphi(x')]=0$ for $x\neq x'$.
In other words, moving a boson through other bosons must not produce
any change of sign. However, moving a fermion through an odd number
of fermions produces a minus sign. This minus sign can be compensated
by a phase factor through a Jordan-Wigner transformation \cite{AltlandSimons2010Book},
\begin{equation}
\varphi(x)=\psi(x)e^{\mathrm{i}\pi\int_{-\infty}^{x}\mathrm{d}y\,\bar{\psi}(y)\psi(y)}\label{eq:Jordan-Wigner_transformation}
\end{equation}
The integral at the exponent, $\int_{-\infty}^{x}\mathrm{d}y\,\bar{\psi}(y)\psi(y)=\int_{-\infty}^{x}\mathrm{d}y\,\sum_{j\in\textrm{particles}}\delta(y-y_{j})$,
counts the number of particles at the left of position $x$. Therefore,
the minus sign coming from moving a fermion through an odd number
of fermions is compensated by the one coming from the phase factor.
Substituting Eq. (\ref{eq:Jordan-Wigner_transformation}) into partition
function (\ref{eq:Partition_Function_Hard-Core_Bosons}) and using
the Grassmann variable identities, $\psi^{2}=\bar{\psi}^{2}=0$, we
obtain
\[
Z=\int\mathcal{D}[\bar{\psi},\psi]e^{\mathrm{i}\int\mathrm{d}t\mathrm{d}x\,\left[\bar{\psi}\left(\mathrm{i}\partial_{t}+\frac{1}{2m}\partial_{x}^{2}+\mu\right)\psi\right]}
\]
This shows that a system of strongly-interacting or hard-core bosons
can be mapped into a system of free fermions. As a consequence, the
chemical potential corresponds to the ground state energy of a system
of fermions with density $n$, that is, $\mu=(\pi n)^{2}/2m=\varepsilon_{\mathrm{F}}$.
It follows that the Tonks-Girardeau gas of hard-core bosons is mapped
into the Tomonaga-Luttinger liquid with,
\begin{equation}
\begin{aligned} & c=v_{\mathrm{F}}=\frac{\pi n}{m}\,,\\
 & K=1\,.
\end{aligned}
\label{eq:Luttinger_parameters_strongly_interacting_bosons}
\end{equation}

\section{Refermionization\label{sec:Refermionization}}

So far, we have derived the low-energy hydrodynamic theory, Eq. (\ref{eq:Luttinger-liquid_Hamiltonian_plus_Corrections}),
starting from interacting fermionic or bosonic particles. In the case
of weakly interacting fermions, the speed of sound, $c$, and the
Luttinger parameter, $K$, are given by Eqs. (\ref{eq:Luttinger_parameters_Fermions})
and the coefficient characterising the cubic density non-linearity
is $\alpha=\pi^{2}/m$. In the case of weakly interacting bosons,
$c$ and $K$ are given by Eqs. (\ref{eq:Luttinger_parameters_Weakly_interacting_bosons})
and $\alpha=0$. Instead, we saw that strongly interacting bosons
behave like free fermions, leading to $c$ and $K$ given by Eqs.
(\ref{eq:Luttinger_parameters_strongly_interacting_bosons}) and $\alpha=\pi^{2}/m$.
Corresponding to the quadratic part of hydrodynamic Hamiltonian (\ref{eq:Luttinger-liquid_Hamiltonian_plus_Corrections})
are action (\ref{eq:Fermion_Bosonization_Linear_Density_Phase}) for
fermions and the quadratic part of action (\ref{eq:Bosonized_boson_Action})
for bosons. These actions are not diagonal and it would be interesting
to know what the excitations of the systems are by diagonalising them.
First, we rescale the phase fields $\theta$ and $\phi$ to the effective
phase fields $\tilde{\theta}$ and $\tilde{\phi}$ as $\theta=\sqrt{K}\,\tilde{\theta}$
and $\phi=\tilde{\phi}/\sqrt{K}$ and the quadratic part of the action
becomes,\label{sec:Phase2EffectiveChiralFields}
\[
S_{0}[\tilde{\theta}^{\alpha},\tilde{\phi}^{\alpha}]=\int\mathrm{d}r\,\left[\left(\begin{array}{c}
\tilde{\theta}^{\alpha}\\
\tilde{\phi}^{\alpha}
\end{array}\right)^{\mathrm{T}}\left(\begin{array}{cc}
\frac{c}{\pi}\partial_{x}^{2} & \frac{1}{\pi}\partial_{t}\partial_{x}\\
\frac{1}{\pi}\partial_{t}\partial_{x} & \frac{c}{\pi}\partial_{x}^{2}
\end{array}\right)\left(\begin{array}{c}
\tilde{\theta}_{\alpha}\\
\tilde{\phi}_{\alpha}
\end{array}\right)-\left(\begin{array}{c}
\frac{2\sqrt{K}}{\pi}\partial_{x}\tilde{\theta}^{\alpha}\\
\frac{2}{\pi\sqrt{K}}\partial_{x}\tilde{\phi}^{\alpha}
\end{array}\right)\left(\begin{array}{c}
u_{\theta\alpha}\\
u_{\phi\alpha}
\end{array}\right)\right]\,.
\]
This action is equal in form to the hydrodynamic action of non-interacting
fermions with a renormalised Fermi velocity $c$ and a rescaled coupling
to the source fields. Since the action of non-interacting fermions
is diagonal in the chiral fields representation, the action is diagonalised
by a transformation to the effective chiral fields $\tilde{\chi}_{\pm}=\tilde{\theta}\pm\tilde{\phi}$,
\[
S_{0}[\tilde{\chi}_{\eta}^{\alpha}]=\int\mathrm{d}r\,\left[\frac{1}{2}\tilde{\chi}_{\eta}^{\alpha}\tilde{D}_{0,\eta}^{-1}\tilde{\chi}_{\eta\alpha}-\frac{2\sqrt{K}}{\pi}\partial_{x}(\tilde{\chi}_{+}^{\alpha}+\tilde{\chi}_{-}^{\alpha})u_{\theta\alpha}-\frac{2}{\pi\sqrt{K}}\partial_{x}(\tilde{\chi}_{+}^{\alpha}-\tilde{\chi}_{-}^{\alpha})u_{\phi\alpha}\right]\,,
\]
where,
\begin{equation}
\tilde{D}_{0,\eta}^{-1}(r)=\frac{\eta}{\pi}\partial_{x}(\partial_{t}+\eta c\partial_{x})\,,\label{eq:Effective_Phonon_Propagator}
\end{equation}
is the inverse propagator (\ref{eq:Chi_Propagator_no_g}) with a renormalised
sound velocity, $c$. Now we turn to the non-linear part of the action.

The non-linear terms are different for fermions and bosons. We start
with fermions. In terms of the rescaled fields $\tilde{\theta}$ and
$\tilde{\phi}$, the non-linear part of the fermionic hydrodynamic
action, Eq. (\ref{eq:Fermion_Bosonization_Non-Linear_Density_Phase}),
becomes,
\[
S_{\mathrm{nl}}[\tilde{\theta}^{\alpha},\tilde{\phi}^{\alpha}]=-\frac{1}{\pi m}\int\mathrm{d}r\,\left\{ \left[K(\tilde{\theta}^{\prime\mathrm{cl}})^{2}+\frac{1}{K}(\tilde{\phi}^{\prime\mathrm{cl}})^{2}\right]\sqrt{K}\tilde{\theta}^{\prime\mathrm{q}}+\frac{2}{\sqrt{K}}\tilde{\theta}^{\prime\mathrm{cl}}\tilde{\phi}^{\prime\mathrm{cl}}\tilde{\phi}^{\prime\mathrm{q}}\right\} \,,
\]
and in terms of the effective chiral fields, $\tilde{\chi}_{\pm}$,
we have,
\begin{equation}
\begin{aligned}S_{\mathrm{nl}}[\tilde{\chi}_{\eta}^{\alpha}] & =-\frac{1}{2\pi m^{*}}\int\mathrm{d}r\,(\tilde{\chi}_{\eta}^{\prime\mathrm{cl}})^{2}\tilde{\chi}_{\eta}^{\prime\mathrm{q}}\\
 & +\frac{1}{8\pi m}\frac{1}{\sqrt{K}}\left(1-K^{2}\right)\int\mathrm{d}r\,\left[(\tilde{\chi}_{\bar{\eta}}^{\prime\mathrm{cl}})^{2}+2\tilde{\chi}_{\bar{\eta}}^{\prime\mathrm{cl}}\tilde{\chi}_{\eta}^{\prime\mathrm{cl}}\right]\tilde{\chi}_{\eta}^{\prime\mathrm{q}}\,,
\end{aligned}
\label{eq:Non-linear_effective_action_fermions}
\end{equation}
where $\bar{\eta}$ is the opposite of $\eta$ and we defined the
effective mass,
\begin{equation}
m^{*}=m\frac{4\sqrt{K}}{3+K^{2}}\approx m\left(1-\frac{3}{8}\delta K^{2}\right)\label{eq:Effective_mass_fermions}
\end{equation}
where we expanded the Luttinger parameter in the weak interaction
limit, $K=1+\delta K$ with $\delta K\ll1$.

In the case of weakly interacting bosons, the non-linear part in Eq.
(\ref{eq:Bosonized_boson_Action}) does not have the density-cube
term, $\sim\theta'^{3}$, present in the fermionic one, Eq. (\ref{eq:Fermion_Bosonization_Non-Linear_Density_Phase}).
Rescaling the fields, Eq. (\ref{eq:Bosonized_boson_Action}) becomes,
\[
S_{\mathrm{nl}}[\tilde{\theta}^{\alpha},\tilde{\phi}^{\alpha}]=-\frac{1}{\pi\sqrt{K}m}\int\mathrm{d}r\,\left\{ (\tilde{\phi}^{\prime\mathrm{cl}})^{2}\tilde{\theta}^{\prime\mathrm{q}}+2\tilde{\theta}^{\prime\mathrm{cl}}\tilde{\phi}^{\prime\mathrm{cl}}\tilde{\phi}^{\prime\mathrm{q}}\right\} \,,
\]
and in terms of $\tilde{\chi}_{\pm}$ we have,
\begin{equation}
\begin{aligned}S_{\mathrm{nl}}[\tilde{\chi}_{\eta}^{\alpha}] & =-\frac{1}{2\pi m^{*}}\int\mathrm{d}r\,(\tilde{\chi}_{\eta}^{\prime\mathrm{cl}})^{2}\tilde{\chi}_{\eta}^{\prime\mathrm{q}}\\
 & +\frac{1}{8\pi m}\frac{1}{\sqrt{K}}\int\mathrm{d}r\,\left[(\tilde{\chi}_{\bar{\eta}}^{\prime\mathrm{cl}})^{2}+2\tilde{\chi}_{\bar{\eta}}^{\prime\mathrm{cl}}\tilde{\chi}_{\eta}^{\prime\mathrm{cl}}\right]\tilde{\chi}_{\eta}^{\prime\mathrm{q}}
\end{aligned}
\label{eq:Non-linear_effective_action_bosons}
\end{equation}
where we defined the effective mass,
\begin{equation}
m^{*}=\frac{4}{3}m\sqrt{K}\,,\label{eq:Effective_mass_bosons}
\end{equation}
and we remind that for weakly interacting bosons $K\gg1$.

In the non-linear actions (\ref{eq:Non-linear_effective_action_fermions})
and (\ref{eq:Non-linear_effective_action_bosons}) the first line
describes interactions between effective chiral fields moving in the
same directions, either $\eta=+1$ or $\eta=-1$; instead, the second
line describes mixed interactions between right, $\eta=+1$, and left,
$\eta=-1$, effective chiral fields. Since the interaction time between
chiral fields moving in opposite directions is negligible compared
to that of those moving in the same direction, we neglect the second
lines of Eqs. (\ref{eq:Non-linear_effective_action_fermions}) and
(\ref{eq:Non-linear_effective_action_bosons}). Then, the hydrodynamic
partition functions for fermionic and bosonic particles have the same
form,
\begin{equation}
Z[u]=\int\mathcal{D}\tilde{\chi}_{\pm}e^{\mathrm{i}\int\mathrm{d}r\,\left[\frac{1}{2}\tilde{\chi}_{\eta}^{\alpha}\tilde{D}_{0,\eta}^{-1}\tilde{\chi}_{\eta\alpha}-\frac{1}{2\pi m^{*}}\left(\tilde{\chi}_{\eta}^{\prime\mathrm{cl}}\right)^{2}\tilde{\chi}_{\eta}^{\prime\mathrm{q}}-\frac{2\sqrt{K}}{\pi}\partial_{x}(\tilde{\chi}_{+}^{\alpha}+\tilde{\chi}_{-}^{\alpha})u_{\theta\alpha}-\frac{2}{\pi\sqrt{K}}\partial_{x}(\tilde{\chi}_{+}^{\alpha}-\tilde{\chi}_{-}^{\alpha})u_{\phi\alpha}\right]}\,.\label{eq:Partition_function_effective_bosonization}
\end{equation}
Integrating over the quantum component of right and left chiral field
\cite{Kamenev2011}, we obtain the equations of motion for the classical
component,
\begin{equation}
(\partial_{t}+\eta c\partial_{x})\tilde{\chi}_{\eta}^{\mathrm{cl}}=-\frac{\eta}{2m^{*}}(\tilde{\chi}_{\eta}^{\mathrm{cl}})^{2}\,,\label{eq:Phonons_Equation_of_Motion}
\end{equation}
where we omitted the source fields.

Partition function (\ref{eq:Partition_function_effective_bosonization})
looks like the partition function of bosonized free fermions with
effective mass $m^{*}$ and a different coupling to density and phase
source fields $u_{\theta}$ and $u_{\phi}$. Comparing the transformation
between fermionic fields and chiral fields, Eq. (\ref{eq:Movers_Chiral_relation}),
we transform to the effective fermionic fields $\bar{\tilde{\psi}}$
and $\tilde{\psi}$ as,$^{\textrm{\ref{fn:KeldyshFactor2}}}$
\[
\frac{1}{2\pi}\partial_{x}\tilde{\chi}_{\eta}\longleftrightarrow\bar{\tilde{\psi}}_{\eta}\tilde{\psi}_{\eta}\,.
\]
Then, the effective chiral fields partition function, Eq. (\ref{eq:Partition_function_effective_bosonization}),
corresponds to the effective fermions partition function,
\begin{equation}
Z[u]=\int\mathcal{D}\bigl[\bar{\tilde{\psi}}_{\pm},\tilde{\psi}_{\pm}\bigr]e^{\mathrm{i}\int_{\rightleftarrows}\mathrm{d}r\,\left[\bar{\tilde{\psi}}_{\eta}(r)\tilde{G}_{\eta}^{-1}(r)\tilde{\psi}_{\eta}(r)-\sqrt{K}(\bar{\tilde{\psi}}_{+}\tilde{\psi}_{+}+\bar{\tilde{\psi}}_{-}\tilde{\psi}_{-})u_{\theta\alpha}-\frac{1}{\sqrt{K}}(\bar{\tilde{\psi}}_{+}\tilde{\psi}_{+}-\bar{\tilde{\psi}}_{-}\tilde{\psi}_{-})u_{\phi\alpha}\right]}\,,\label{eq:Left-Right_effective_movers_partition_function}
\end{equation}
where the Green's functions is,
\begin{equation}
\tilde{G}_{\eta}^{-1}(r)=\mathrm{i}\partial_{t}+\eta\mathrm{i}c\partial_{x}+\frac{\partial_{x}^{2}}{2m^{*}}\,,\label{eq:Right-Left_movers_inverse_Green_Function-1}
\end{equation}
which leads to the spectrum,
\begin{equation}
\varepsilon_{\eta}(k)=\eta ck+\frac{k^{2}}{2m^{*}}\,.\label{eq:Fermion_quasiparticle_spectrum}
\end{equation}
 The coupling to the source fields tells us that density and phase
in terms of the effective fermions are $\sqrt{K}(\bar{\tilde{\psi}}_{+}\tilde{\psi}_{+}+\bar{\tilde{\psi}}_{-}\tilde{\psi}_{-})$
and $(\bar{\tilde{\psi}}_{+}\tilde{\psi}_{+}-\bar{\tilde{\psi}}_{-}\tilde{\psi}_{-})/\sqrt{K}$.
These relations together with Eqs. (\ref{eq:Left-Right_effective_movers_partition_function})
and (\ref{eq:Right-Left_movers_inverse_Green_Function-1}) are known
as refermionization. We do not prove it rigorously here and we refer
to Refs. \cite{Rozhkov2005,Rozhkov2008,DelftSchoeller1998} for a
rigorous derivation.

As in the case of bosonization, for which the non-linearity of the
spectrum must be small, $k\ll mv_{\mathrm{F}}$, for refermionization
to be possible the condition $k\ll m^{*}c$ must hold. An additional
condition in the case of weakly interacting bosons requires $p\ll p_{\mathrm{G}}$,
where $p_{\mathrm{G}}=4(m/m^{*})mc=3mc/\sqrt{K}$ is the Ginzburg
momentum \cite{PustilnikMatveev2014,Ristivojevic2014}. The reason
is that, only for momenta smaller than $p_{\mathrm{G}}$ the spectrum
of weekly interacting bosons is quadratic \cite{PustilnikMatveev2014,Ristivojevic2014,KulishManakovFaddeev76},
which is one of requirement for refermionization. The Ginzburg momentum
is found by comparing the value of the momentum for which the quadratic
spectrum,
\[
\varepsilon(p)=cp+\frac{p^{2}}{2m^{*}}\,,\quad p\ll p_{\mathrm{G}}\,,
\]
crosses over to the Bogoliubov spectrum,
\[
\varepsilon(p)=cp\sqrt{1+\frac{p^{2}}{4m^{2}c^{2}}}\approx cp+\frac{p^{3}}{8m^{2}c}\,,\quad p\gg p_{\mathrm{G}}\,.
\]
The two energies match at momentum $p_{\mathrm{G}}$, where we used
Eq. (\ref{eq:Effective_mass_bosons}) for the value of the effective
mass. As $K\gg1$ for weakly interacting bosons, the condition for
the validity of refermionization is restrictive because $p_{\mathrm{G}}\ll mc$.
Finally, we note that fermionic quasiparticles are lighter than the
initial particles, $m^{*}<m$, for interacting fermions and heavier,
$m^{*}>m$, for weakly interacting bosons.

\section{Particle-excitation map}

\begin{figure}[t]
\centering
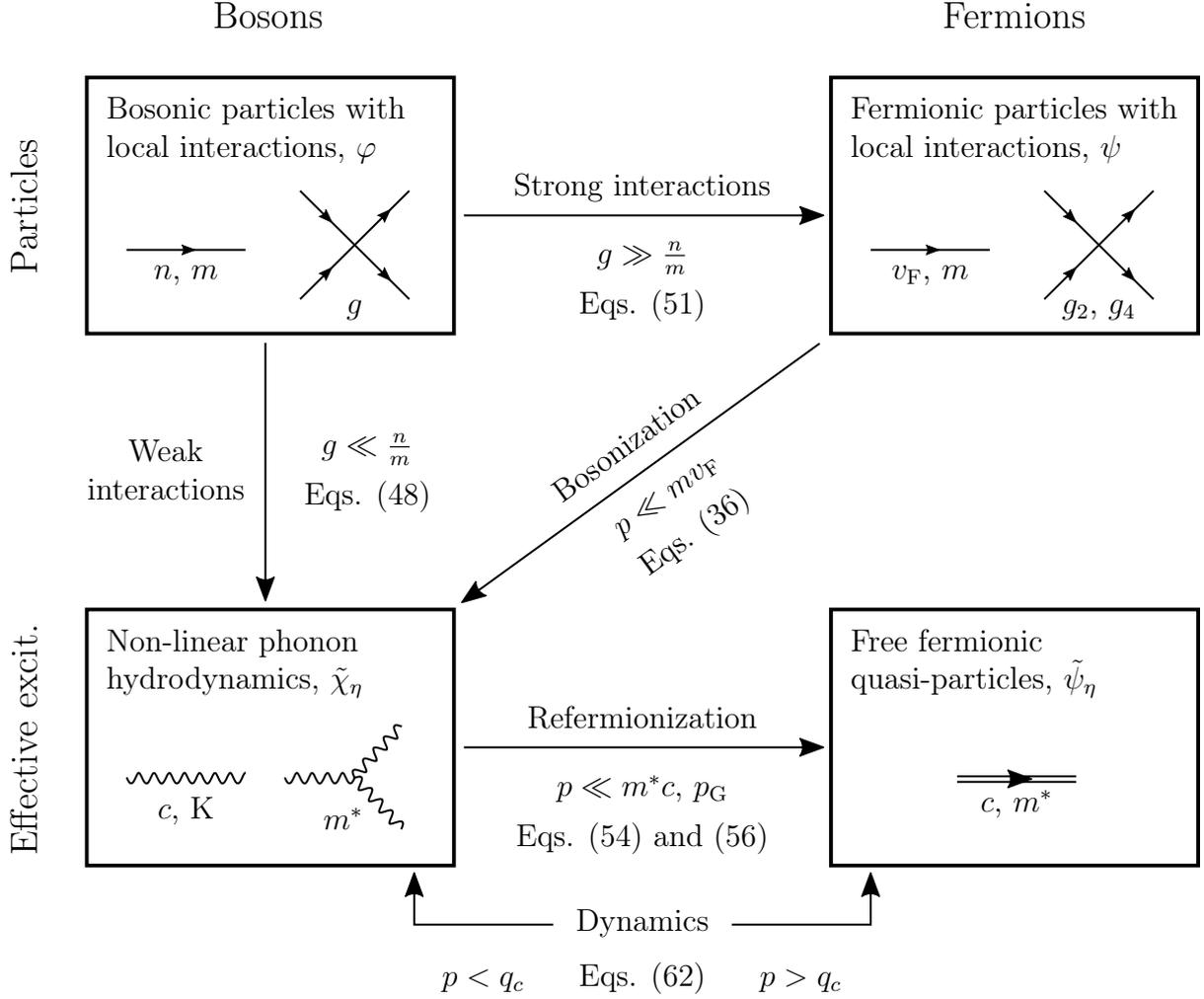

\caption[Schematic summary of the main results of this work \ref{chap:One-dimensional-systems}.]{Schematic summary of the main results of this work. Left and right
columns refer to bosonic and fermionic statistics and top and bottom
rows to particles and effective excitations. The equation references
give the relation between the parameters of the related theories.
\label{fig:Bosonization_Fermionization_scheme}}
\end{figure}

In this work we derived the low-energy hydrodynamic and the fermionic
quasiparticle theories for interacting fermionic and bosonic particles.
The relations between these theories are depicted in Fig. \ref{fig:Bosonization_Fermionization_scheme}
with the help of Feynman diagrams.
 We started from fermions, the top-right box in Fig. \ref{fig:Bosonization_Fermionization_scheme}.
The fermionic theory is characterised by the propagator and the local
density-density interaction term, represented by the straight line
and the cross diagrams. The propagator depends on the Fermi velocity,
$v_{\mathrm{F}}$, and mass, $m$, and the interaction terms correspond
to scattering with strengths $g_{2}$ and $g_{4}$ between two fermions.
Using functional bosonization, we showed that the low-energy particle-hole
excitations around the two Fermi points give rise to phononic excitations.
The theory of phononic excitations is characterised by the two diagrams
in the bottom-left box in Fig. \ref{fig:Bosonization_Fermionization_scheme}.
The wiggly line is the phonon propagator that depends on the speed
of sound, $c$, and the Luttinger parameter, $K$. The three-leg diagram
represents the non-linear contribution proportional to $m^{*}$ that
comes from the mass curvature, $m$, of the fermionic spectrum and
corresponds to the three-loop of Fig. \ref{fig:Vertex_Corrections}.
This non-linear term, as suggested by the diagram, corresponds to
the processes of splitting a phonon into two or recombining two into
one. Additional non-linear terms with more legs are present, but can
be neglected for small momenta, $p\ll mv_{\mathrm{F}}$. Then, we
considered bosonic particles with contact interactions, the top-left
box in Fig. \ref{fig:Bosonization_Fermionization_scheme}. The bosonic
theory is characterised by the propagator and the contact density-density
interaction term, represented by the straight line and the cross diagrams.
The propagator depends on the density, $n$, and mass, $m$, and the
interaction term corresponds to the scattering with strength $g$
between two bosons. We considered the two limits of weak and strong
interactions. In the case of weak interactions, $g\ll\frac{n}{m}$,
the hydrodynamic theory was derived in terms of density and phase
fluctuations over the average values. In the case of strong interactions,
$g\gg\frac{n}{m}$, bosons become impenetrable, a feature that allows
the mapping to free fermions and, in turn, to the hydrodynamic theory
with $K=1$. Finally, we mapped the non-linear hydrodynamic theory
to free fermionic quasiparticles with an effective Fermi velocity
$c$ and an effective mass $m^{*}$, the bottom-right box in Fig.
\ref{fig:Bosonization_Fermionization_scheme}. The mapping requires
the additional low-momentum condition $p\ll m^{*}c$ and in the case
of weakly interacting bosons the additional condition $p\ll p_{\mathrm{G}}\sim1/\sqrt{K}$.

\section{Dynamics at finite temperatures}

Non-linear phonons and free fermionic quasiparticles are equally good
at describing static properties \cite{ImambekovGlazman2012}. However,
this is not the case for dynamical ones. In fact, to realise this,
it is sufficient to compare the dynamical structure factor, $S(q,\omega)$,
calculated using the two representations. We have already calculated
$S(q,\omega)$ using free fermions and the result is represented in
Fig. \ref{fig:DSF}. In the case of fermionic quasiparticles, Fermi
velocity, $v_{\mathrm{F}}$, and mass, $m$, are replaced by $c$
and $m^{*}$ but the width depends on $q$ in the same way: $\delta\omega(q)\sim q^{2}$
at $T=0$ and $\delta\omega_{T}(q)\sim q$ at $T>0.$ Using phonons,
we start deriving the dynamical structure factor using the quadratic
approximation. We do not need to do the calculation explicitly, we
just need to know that the dynamical structure factor is non-zero
where the spectrum of particle-holes excitations is non-zero. Because
the quadratic phonons approximation corresponds to fermions with linearised
spectrum, there is no spectrum curvature and particle-hole excitations
can only satisfy the relation $\omega=cq$. It follows that the dynamical
structure factor is a straight line with slope $c$ and no width in
the $q-\omega$ plane of Fig. \ref{fig:DSF}. This result is in clear
contrast with the one derived using fermionic quasiparticles. But
we know that, when there are no interactions between fermions, particles
and quasiparticles coincide, giving the correct result. Then, we may
add the three-leg term to the quadratic phonons. With this term, the
calculation of the width of the dynamical structure factor is not
straightforward as it involves a self-consistent approach to avoid
a resonant behaviour. We do not give the explicit calculation here
and refer to Refs. \cite{Andreev1980,Samokhin1998}. The resulting
width is $\delta\omega(q)\sim q^{2}$ at $T=0$ and $\delta\omega_{T}(q)\sim q^{3/2}$
at $T>0.$ Now, fermionic quasiparticles and phonons give the same
result at $T=0$ but differ at $T>0$: even with the three-leg correction,
the two representations lead to different thermal dynamics in the
small-momentum limit, where the three-leg approximation should hold
better. This leads to an apparent paradox: which one is the best representation
to describe the thermal dynamics of one-dimensional systems, non-linear
phonons or free fermionic quasiparticles? This question is answered
in Refs. \cite{ABG2014,BovoPhDThesis} through the study of the dynamical
structure factor, which shows that, at momenta lower than
\begin{equation}
q_{c}=\frac{\pi^{5}}{128}[\Gamma_{0}']^{2}\frac{T^{7}}{m^{*2}c^{7}}\,,\label{eq:Thermal_Dynamics_Time}
\end{equation}
the dynamics is dominated by interacting phonons and, at momenta higher
that $q_{c}$, the dynamics is dominated by free fermionic quasiparticles.
Here, $\Gamma_{0}'$ is a parameter that depends on the specific details
of the systems.

\section{Conclusions}

In this work we gave a self-contained presentation of non-linear bosonization
and refermionization of one-dimensional quantum systems within the
Keldysh functional integral. We started with the derivation of the
Tomonaga-Luttinger liquid for a system of interacting fermions by
considering the linear spectrum approximation, that amounts to neglecting
the spectrum curvature. This is a good approximation at low energies,
where there are mainly low-energy particle-hole excitations around
the two Fermi points and works well for static systems. In order to
study dynamical systems, where higher energies become important, we
derived an infinite series of non-linear corrections arising from
the fermionic spectrum curvature and calculated the first one, corresponding
to the interaction between three phonons. The result, obtained within
the Keldysh formalism, was checked against a more rigorous approach
based on the Matsubara formalism, presented in the Appendix, by evaluating
the low-energy asymptotic contribution of the three- and four-phonon
interactions to the action and some properties of the five-phonon
interaction. These results and dimensional analysis invited a conjecture
on the asymptotic curvature and temperature dependence of the interaction
terms between an arbitrary number of phonons.

To complement the bosonization of interacting fermions, we bosonized
interacting bosonic particles in the weak and strong interaction limits.
In the weak-interaction limit, the standard mean-field approach was
sufficient to derive the effective phononic action, but the strong-interaction
limit required a more creative solution by decoupling the interaction
term with a Hubbard-Stratonovich transformation and showing how a
system of bosons with infinite repulsive interaction can be restated
as a system of free fermions.

Finally, we used the relation between non-linear phonons and interacting
fermions to refermionize the system, that is, to derive an approximate
theory of fermionic quasiparticles. This work culminates in a map
between fermionic and bosonic particles and bosonic and fermionic
excitations, that is, phonons and fermionic quasiparticles. In particular,
we noticed that the description of one-dimensional quantum systems
in term of phonons and fermionic quasiparticles should be equivalent.
However, the dynamical structure factor shows that, within the approximations
that we considered, the equivalence applies only to static and zero-temperature
dynamical properties and leads to different results for the thermal
dynamics, where phonons and fermionic quasiparticles are better at
describing respectively lower and higher momenta.

\section{Acknowledgements}

I am grateful to D. M. Gangardt, A. J. Kingl and M. Jones for helpful
discussions. I acknowledge the support of the University of Birmingham.

\appendix

\section{n-loop reduction formula\label{sec:N-loop-reduction-formula}}

\begin{figure}
\centering
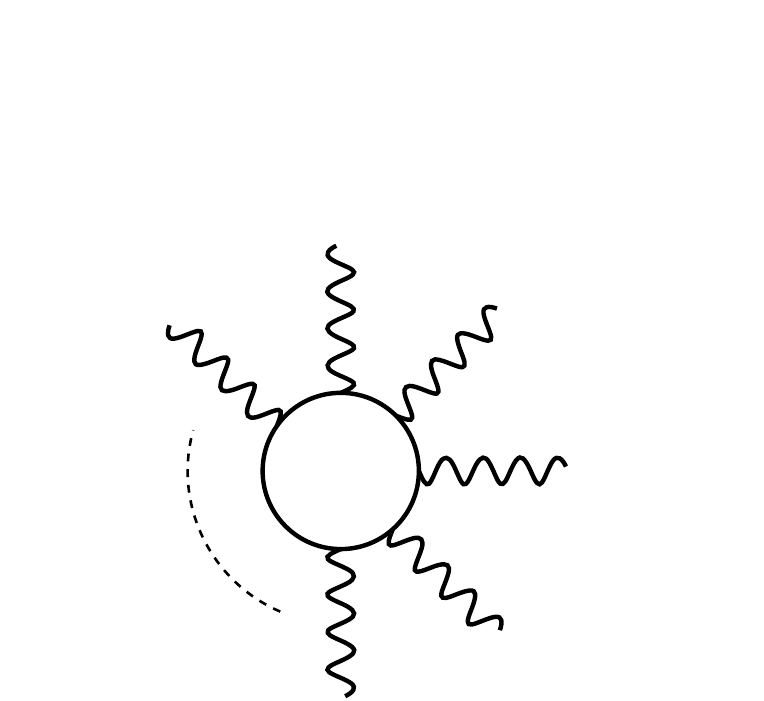

\caption{n-loop.\label{fig:nloop}}
\end{figure}

\begin{figure}
\centering
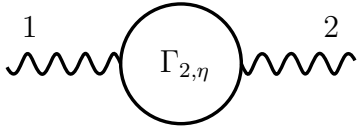

\caption{Two-loop.\label{fig:2loop}}
\end{figure}

In this section we derive a reduction formula for the symmetrised
fermionic n-loop by adapting the zero temperature derivation of Refs.
\cite{NeumayrMetzner1998,NeumayrMetzner1999} to finite temperatures,
$T$, using the Matsubara formalism. The fermionic n-loop is shown
in Fig. \ref{fig:nloop} and is defined as (compare it with the analogous
expression (\ref{eq:n-vertex}) in the Keldysh formalism),
\[
\Gamma_{n}(Q_{1},\ldots,Q_{n-1})=I_{n}(P_{1},\ldots,P_{n})=-\frac{1}{\beta}\sum_{E_{\ell}}\int\frac{\mathrm{d}k}{2\pi}G(K+P_{1})\ldots G(K+P_{n})\,,
\]
where $Q_{j}=(\omega_{j},q_{j})$ are phononic momentum and Matsubara
energy variables and $P_{j}=(\epsilon_{j},p_{j})=Q_{1}+\ldots+Q_{j}$
and $K=(E_{\ell},k)$ are fermionic ones. The Matsubara Green's function
is defined as,
\[
G(K)=\frac{1}{\mathrm{i}E_{\ell}-\varepsilon(k)}\,,
\]
and $\varepsilon(k)=v_{\mathrm{F}}k+k^{2}/2m$ is the right mover
spectrum. Here we limit the study to right movers as the results for
left movers are symmetric. The sum over fermionic energies $E_{\ell}=\pi\mathrm{i}(2\ell+1)T$
is easily evaluated by means of the residue theorem and yields,
\[
I_{n}(P_{1},\ldots,P_{n})=\int\frac{\mathrm{d}k}{2\pi}\sum_{i=1}^{n}n_{i}\prod_{\substack{j=1\\
j\neq i
}
}^{n}\frac{1}{f_{ji}(k)}\,,
\]
where,
\[
n_{i}=n_{\mathrm{F}}(-\mathrm{i}\epsilon_{i}+\varepsilon(k+p_{i}))\,,
\]
is the occupation factor of a state with momentum $p_{i}$ and energy
$\epsilon_{i}$ and,
\[
\begin{aligned}f_{ij}(k) & =(\mathrm{i}\epsilon_{i}-\varepsilon(k+p_{i}))-(\mathrm{i}\epsilon_{j}-\varepsilon(k+p_{j}))\\
 & =(\mathrm{i}\epsilon_{i}-\varepsilon(p_{i}))-(\mathrm{i}\epsilon_{j}-\varepsilon(p_{j}))-\frac{p_{i}-p_{j}}{m}k\,.
\end{aligned}
\]
The n-loop can be rewritten as,
\begin{equation}
I_{n}(P_{1},\ldots,P_{n})=\int\frac{\mathrm{d}k}{2\pi}\sum_{i=1}^{n}n_{i}\prod_{\substack{j=1\\
j\neq i
}
}^{n}\left(\frac{m}{p_{ij}}\right)\prod_{\substack{j=1\\
j\neq i
}
}^{n}\left(\frac{1}{k-\Delta_{ij}}\right)\,,\label{eq:n-loop_Matsubara}
\end{equation}
where $p_{ij}=p_{i}-p_{j}$ and,
\[
\Delta_{ij}=\frac{m}{p_{ij}}[(\mathrm{i}\epsilon_{i}-\varepsilon(p_{i}))-(\mathrm{i}\epsilon_{j}-\varepsilon(p_{j}))]\,.
\]
The last product in the n-loop (\ref{eq:n-loop_Matsubara}) can be
expressed as a sum using a partial fraction expansion,
\[
\prod_{\substack{j=1\\
j\neq i
}
}^{n}\left(\frac{1}{k-\Delta_{ij}}\right)=\sum_{\substack{j=1\\
j\neq i
}
}^{n}\left[\frac{1}{k-\Delta_{ij}}\prod_{\substack{n=1\\
n\neq i,j
}
}^{n}\left(\frac{1}{\Delta_{ij}-\Delta_{in}}\right)\right]\,.
\]
Then, the n-loop (\ref{eq:n-loop_Matsubara}) becomes,
\[
I_{n}(P_{1},\ldots,P_{n})=\sum_{\substack{i,j=1\\
i\neq j
}
}^{n}\left(\frac{m}{p_{ij}}\int\frac{\mathrm{d}k}{2\pi}\frac{n_{i}}{k-\Delta_{ij}}\right)\prod_{\substack{n=1\\
n\neq i,j
}
}^{n}\left(\frac{1}{f_{ij}^{n}}\right)\,,
\]
where,
\[
\begin{aligned}f_{ij}^{n} & =\frac{p_{in}}{m}(\Delta_{ij}-\Delta_{in})=\mathrm{i}\frac{p_{jn}\epsilon_{in}-p_{in}\epsilon_{jn}}{p_{ij}}+\frac{p_{in}p_{jn}}{2m}\\
 & =\frac{1}{p_{ij}}\left[p_{jl}(\tilde{\varepsilon}_{i}-\frac{p_{i}^{2}}{2m})-p_{il}(\tilde{\varepsilon}_{j}-\frac{p_{j}^{2}}{2m})+p_{ij}(\tilde{\varepsilon}_{l}-\frac{p_{l}^{2}}{2m})\right]\,.
\end{aligned}
\]
Using the symmetries $f_{ij}^{n}=f_{ji}^{n}$ and $\Delta_{ij}=\Delta_{ji}$,
we expressed the sum using the difference of occupation factors,
\[
I_{n}(P_{1},\ldots,P_{n})=\sum_{\substack{i,j=1\\
i>j
}
}^{n}\left(\frac{m}{p_{ij}}\int\frac{\mathrm{d}k}{2\pi}\frac{n_{i}-n_{j}}{k-\Delta_{ij}}\right)\prod_{\substack{n=1\\
n\neq i,j
}
}^{n}\left(\frac{1}{f_{ij}^{n}}\right)\,.
\]
The term in the first parenthesis coincides with the two-loop,
\begin{equation}
\begin{aligned}I_{2}(P_{i},P_{j}) & =-\frac{1}{\beta}\sum_{E_{\ell}}\int\frac{\mathrm{d}k}{2\pi}G(K+P_{i})G(K+P_{j})\\
 & =\frac{m}{p_{ij}}\int\frac{\mathrm{d}k}{2\pi}\frac{n_{i}-n_{j}}{k-\Delta_{ij}}\,,
\end{aligned}
\label{eq:Two-loop}
\end{equation}
shown in Fig. \ref{fig:2loop}. Making the shift $K\rightarrow K-P_{j}$,
we obtain,
\[
I_{2}(P_{i},P_{j})=-\frac{1}{\beta}\sum_{E_{\ell}}\int\frac{\mathrm{d}k}{2\pi}G(K+P_{i}-P_{j})G(K)=I_{2}(P_{i}-P_{j},0)\equiv I_{2}(P_{i}-P_{j})\,.
\]
The two-loop depends only on the differences $p_{ij}=p_{i}-p_{j}$
and $\epsilon_{ij}=\epsilon_{i}-\epsilon_{j}$ as a consequence of
energy and momentum conservations. Finally, the reduction formula
becomes,
\begin{equation}
I_{n}(P_{1},\ldots,P_{n})=\sum_{\substack{i,j=1\\
i>j
}
}^{n}I_{2}(P_{ij})\prod_{\substack{n=1\\
n\neq i,j
}
}^{n}\left(\frac{1}{f_{ij}^{n}}\right)\,.\label{eq:n-loop_reduction_formula}
\end{equation}

\subsection{Two-loop}

We rewrite the two-loop (\ref{eq:Two-loop}) as, 
\[
I_{2}(P_{i},P_{j})=-\frac{m}{4\pi p_{ij}}\int_{-\infty}^{\infty}\mathrm{d}k\frac{\tanh\left(\frac{\varepsilon(k+p_{i})}{2T}\right)-\tanh\left(\frac{\varepsilon(k+p_{j})}{2T}\right)}{k-\Delta_{ij}}\,,
\]
In this form, a shift of the variable $k$ cannot be taken separately
for the two hyperbolic tangents, as only their difference is convergent
at $k\rightarrow\pm\infty$. To allow for the shift to be taken for
the two terms separately, we integrate by parts,
\[
I_{2}(P_{i},P_{j})=\frac{m}{8\pi Tp_{ij}}\int\mathrm{d}k\log\left(k-\Delta_{ij}\right)\left[\frac{v_{\mathrm{F}}+(k+p_{i})/m}{\cosh^{2}\left(\frac{\varepsilon(k+p_{i})}{2T}\right)}-\frac{v_{\mathrm{F}}+(k+p_{j})/m}{\cosh^{2}\left(\frac{\varepsilon(k+p_{j})}{2T}\right)}\right]\,.
\]
Now the two terms converge separately and we make the shifts $k\rightarrow k-p_{i}$
in the first one and $k\rightarrow k-p_{j}$ in the second one to
obtain,
\begin{equation}
\begin{aligned}I_{2}(P_{ij}) & =\frac{m}{8\pi Tp_{ij}}\int\mathrm{d}k\log\left(\frac{k-p_{i}-\Delta_{ij}}{k-p_{j}-\Delta_{ij}}\right)\frac{v_{\mathrm{F}}+k/m}{\cosh^{2}\left(\frac{\varepsilon(k)}{2T}\right)}\\
 & =\frac{m}{8\pi Tp_{ij}}\int\mathrm{d}k\log\left[\frac{\tilde{\epsilon}_{ij}-\frac{p_{ij}}{m}(k-\frac{p_{ij}}{2})}{\tilde{\epsilon}_{ij}-\frac{p_{ij}}{m}(k+\frac{p_{ij}}{2})}\right]\frac{v_{\mathrm{F}}+k/m}{\cosh^{2}\left(\frac{v_{\mathrm{F}}k+k^{2}/2m}{2T}\right)}\,,
\end{aligned}
\label{eq:2Loop_Finite_Temp}
\end{equation}
where we defined $\tilde{\epsilon}_{ij}=\mathrm{i}\epsilon_{ij}-v_{\mathrm{F}}p_{ij}$,
where the double index from now on denotes the difference, for example
$\tilde{\epsilon}_{ij}=\tilde{\epsilon}_{i}-\tilde{\epsilon}_{j}$.
At zero temperature,
\[
\lim_{T\rightarrow0}\frac{\text{1}}{4T\cosh^{2}\left(\frac{v_{\mathrm{F}}k+k^{2}/2m}{2T}\right)}=\delta\left(v_{\mathrm{F}}k+\frac{k^{2}}{2m}\right)\,.
\]
The zeros of the argument of the delta function are at $k=0$ and
$k=-2mv_{\mathrm{F}}=-2k_{F}$, but we can neglect the second zero
as it describes a sub-leading contribution with momenta close to the
the left Fermi point. Integrating over $k$, the two-loop simplifies
to,
\[
I_{2,0}(P_{ij})=\frac{m}{2\pi p_{ij}}\log\left[\frac{\tilde{\epsilon}_{ij}+\frac{p_{ij}^{2}}{2m}}{\tilde{\epsilon}_{ij}-\frac{p_{ij}^{2}}{2m}}\right]\,.
\]
Noting that $I_{2}$ is even in $m$, an expansion in powers of $m^{-1}$
at zero temperature reads,
\[
I_{2,0}(P_{ij})=\frac{m}{\pi p_{ij}}\sum_{\substack{n=1\\
n\textrm{ odd}
}
}^{\infty}\frac{1}{n}\left(\frac{p_{ij}^{2}}{2m\tilde{\epsilon}_{ij}}\right)^{n}=\frac{1}{2\pi}\frac{p_{ij}}{\tilde{\epsilon}_{ij}}+\frac{1}{24\pi m^{2}}\frac{p_{ij}^{5}}{\tilde{\epsilon}_{ij}^{3}}+\mathcal{O}(m^{-4})
\]
The dynamical condition for the convergence of the series is $p_{ij}^{2}\ll m\tilde{\epsilon}_{ij}$.
Instead, in the static limit, $\epsilon_{i}\rightarrow0$, we have
$\tilde{\epsilon}_{ij}\rightarrow-cp_{ij}$ and the condition for
convergence becomes $|p_{ij}|=|q_{i}+\ldots+q_{j-1}|\ll mv_{\mathrm{F}}=k_{\mathrm{F}}$,
which means that the phononic momenta must be small compared to the
Fermi momentum, as seen in Chap. \ref{chap:One-dimensional-systems}.
At finite temperatures, the two-loop gets an additional term dependent
on $T$ in the second term of the expansion in $m^{-1}$,
\[
I_{2}(P_{ij})=\frac{1}{2\pi}\frac{p_{ij}}{\tilde{\epsilon}_{ij}}+\frac{1}{24\pi m^{2}}\frac{p_{ij}^{2}}{\tilde{\epsilon}_{ij}^{3}}\left[p_{ij}^{3}+2\pi^{2}\frac{T^{2}}{c^{3}}(2cp_{ij}-\tilde{\epsilon}_{ij})\right]+\mathcal{O}(m^{-4})\,.
\]
Note that we did not expand in $T$.

\subsection{Three-loop and four-loop}

\begin{figure}
\centering
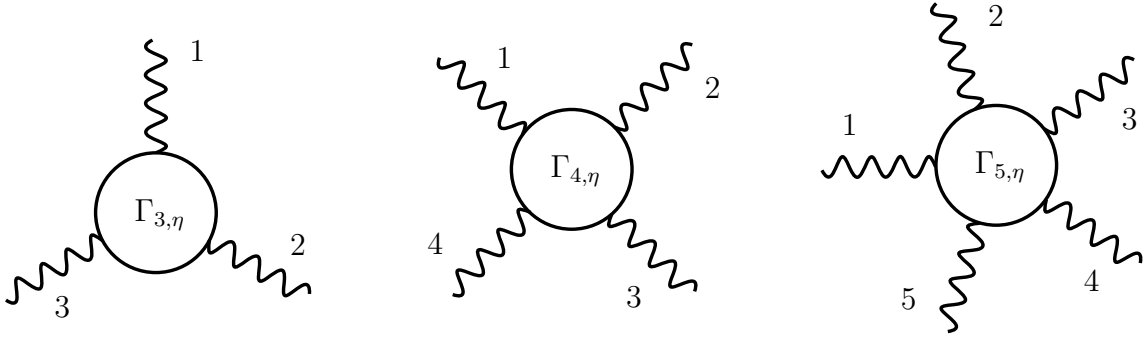

\caption{Three-loop, four-loop and five-loop.\label{fig:345loops}}
\end{figure}

The three-loop, the first in Fig. \ref{fig:345loops}, is given in
terms of $q_{i}$ and $\omega_{i}$, by,
\[
\begin{aligned}\Gamma_{3}(Q_{1},Q_{2}) & =I_{3}(P_{1}=0,P_{2}=Q_{1},P_{3}=Q_{1}+Q_{2})\\
 & =\frac{q_{1}I_{2}(Q_{1})+q_{2}I_{2}(Q_{2})-(q_{1}+q_{2})I_{2}(Q_{1}+Q_{2})}{\mathrm{i}(q_{1}\epsilon_{2}-q_{2}\epsilon_{1})+q_{1}q_{2}(q_{1}+q_{2})/2m}\,.
\end{aligned}
\]
n-loops are symmetric by exchange of external legs. Therefore, we
symmetrise by taking the circular permutations of $Q_{1}$, $Q_{2}$
and $Q_{3}$, which amount to summing $\Gamma_{3}(Q_{1},Q_{2})$ and
$\Gamma_{3}(Q_{2},Q_{1})$. The symmetric part of the three loop is,
\[
\begin{aligned}\widetilde{\Gamma}_{3}(Q_{1},Q_{2}) & =\Gamma_{3}(Q_{1},Q_{2})+\Gamma_{3}(Q_{2},Q_{1})\\
 & =-\frac{q_{1}q_{2}(q_{1}+q_{2})}{2m}\frac{q_{1}I_{2}(Q_{1})+q_{2}I_{2}(Q_{2})-(q_{1}+q_{2})I_{2}(Q_{1}+Q_{2})}{(q_{1}\epsilon_{2}-q_{2}\epsilon_{1})^{2}+(q_{1}q_{2}(q_{1}+q_{2})/2m)^{2}}\,.
\end{aligned}
\]
An $m^{-1}\rightarrow0$ power expansion reads,
\[
\begin{aligned}\widetilde{\Gamma}_{3}(Q_{1},Q_{2}) & =\frac{1}{4\pi m}\frac{q_{1}q_{2}q_{3}}{\tilde{\omega}_{1}\tilde{\omega}_{2}\tilde{\omega}_{3}}\\
 & +\frac{1}{8\pi m^{3}}q_{1}q_{2}q_{3}\left\{ \frac{1}{(q_{2}\tilde{\omega}_{1}-q_{1}\tilde{\omega}_{2})^{2}}\left[\frac{1}{2}\frac{q_{1}^{2}q_{2}^{2}q_{3}^{2}}{\tilde{\omega}_{1}\tilde{\omega}_{2}\tilde{\omega}_{3}}-\frac{1}{3}\left(\frac{q_{1}^{6}}{\tilde{\omega}_{1}^{3}}+\frac{q_{2}^{6}}{\tilde{\omega}_{2}^{3}}+\frac{q_{3}^{6}}{\tilde{\omega}_{3}^{3}}\right)\right.\right.\\
 & \left.\left.-\frac{4}{3}\pi^{2}\frac{T^{2}}{c^{2}}\left(\frac{q_{1}^{4}}{\tilde{\omega}_{1}^{3}}+\frac{q_{2}^{4}}{\tilde{\omega}_{2}^{3}}+\frac{q_{3}^{4}}{\tilde{\omega}_{3}^{3}}\right)\right]-\frac{2}{3}\pi^{2}\frac{T^{2}}{c^{3}}\frac{q_{1}\tilde{\omega}_{2}\tilde{\omega}_{3}+\tilde{\omega}_{1}q_{2}\tilde{\omega}_{3}+\tilde{\omega}_{1}\tilde{\omega}_{2}q_{3}}{\tilde{\omega}_{1}^{2}\tilde{\omega}_{2}^{2}\tilde{\omega}_{3}^{2}}\right\} \\
 & +\mathcal{O}(m^{-5})\,,
\end{aligned}
\]
where $\tilde{\omega}_{i}=\mathrm{i}\omega_{i}-v_{\mathrm{F}}q_{i}$
and $Q_{3}=-Q_{1}-Q_{2}$. We note that at the first order in $m^{-1}$,
the three-loop does not depend on temperature. Moreover, there is
no term independent of $m$, which leads to $\lim_{m^{-1}\rightarrow0}\widetilde{\Gamma}_{3}=0$,
consistent with the Dzyaloshinski-Larking theorem (see page \pageref{subsec:Dzyaloshinski-Larking theorem}).
Similarly, the symmetrised four-loop, the second in Fig. \ref{fig:345loops},
is,
\[
\widetilde{\Gamma}_{4}(Q_{1},Q_{2},Q_{3})=\frac{1}{4\pi m^{2}}\frac{q_{1}q_{2}q_{3}q_{4}}{\tilde{\omega}_{1}\tilde{\omega}_{2}\tilde{\omega}_{3}\tilde{\omega}_{4}}\left[\frac{q_{1}+q_{2}}{\tilde{\omega}_{1}+\tilde{\omega}_{2}}+\frac{q_{1}+q_{3}}{\tilde{\omega}_{1}+\tilde{\omega}_{3}}+\frac{q_{2}+q_{3}}{\tilde{\omega}_{2}+\tilde{\omega}_{3}}\right]+\mathcal{O}(m^{-4})\,.
\]
As for the three-loop, at the first order in $m^{-1}$, the three-loop
does not depend on temperature and it is consistent with the Dzyaloshinski-Larking
theorem, $\lim_{m^{-1}\rightarrow0}\widetilde{\Gamma}_{4}=0$. However,
$\widetilde{\Gamma}_{4}=\mathcal{O}(m^{-2})$ goes to zero faster
than $\widetilde{\Gamma}_{3}=\mathcal{O}(m^{-1})$ for $m^{-1}\rightarrow0$.
Moreover, it can be checked that the five-loop, the third in Fig.
\ref{fig:345loops}, is $\widetilde{\Gamma}_{5}=\mathcal{O}(m^{-3})$
for $m^{-1}\rightarrow0$ and at the lowest order, $m^{-3}$, $\widetilde{\Gamma}_{5}$
does not depend on temperature.\footnote{The symmetrisation was done numerically and was computationally too
expensive for loops with more than five legs. }

\bibliographystyle{unsrt}
\bibliography{Bibliography}

\end{document}